\documentclass[%
 amsmath,amssymb,
 aps,longbibliography
]{revtex4-1}

\usepackage{graphicx}
\usepackage{dcolumn}
\usepackage{bm}
\usepackage{multirow}

\begin{document}

\preprint{APS/123-QED}

\title{Thermal convection of viscoelastic fluids in concentric rotating cylinders: Elastic turbulence and kinetic energy budget analysis}
\author{A. Chauhan}
\author{C. Sasmal}
\email{csasmal@iitrpr.ac.in}
\affiliation{Department of Chemical Engineering, Indian Institute of Technology
Ropar, Rupnagar, Punjab, 140001, India}


\date{\today}

\begin{abstract}
The introduction of solid polymers into a Newtonian solvent induces significant modifications in the flow behavior and heat transfer characteristics of resulting viscoelastic fluids. This study performs a comprehensive numerical investigation on thermal convection within a system comprising two concentric horizontal cylinders filled with viscoelastic fluids, with the inner cylinder rotating. The analysis encompasses all three modes of thermal convection, namely, forced, free, and mixed convection, over a range of Weissenberg numbers up to 10 and three values of the Richardson number, namely, 0, 0.143, and $\infty$, representing forced, mixed, and free convection modes of heat transfer, respectively. In forced convection, the flow field remains stable, while in free and mixed convection, an increase in the Weissenberg number leads to a transition from steady to unsteady periodic, quasi-periodic, and finally, an aperiodic and chaotic behavior. This transition arises due to the presence of elastic instability and the subsequent appearance of elastic turbulence in viscoelastic fluids with the increasing Weissenberg number. Furthermore, our findings indicate that fluid viscoelasticity has minimal influence on heat transfer rates in the cases of forced and free convection. Conversely, heat transfer rates in mixed convection increase with the Weissenberg number. We conduct a detailed analysis of the viscoelastic kinetic energy budget to elucidate this enhancement in the heat transfer rate for viscoelastic fluids. We show that this improved heat transfer results from kinetic energy transfer from polymer molecules to the flow field, leading to increased chaotic motion within the system and, eventually, higher heat transfer rates.
\end{abstract}

\maketitle


\section{\label{sec:1}Introduction}

Over the past several decades, the study of mixed convection heat transfer phenomena within a configuration comprising two horizontal concentric cylinders, known as a concentric annulus, has garnered substantial attention from the scientific community~\cite{gazley1958heat,bjorklund1959heat}. This system often involves the presence of one or both cylinders rotating, which adds an additional layer of complexity to the investigation. This particular configuration finds extensive practical and engineering applications, ranging from heat exchangers and nuclear reactors to thermal solar collectors and power generation systems. The fluid dynamics within this annular geometry become crucial in understanding the heat transfer processes involved~\cite{taylor,di2005instabilities,coles}. The interaction between forced convection induced by the rotation of the inner cylinder and natural convection driven by temperature gradients leads to intriguing phenomena. As a result, numerous researchers have delved into this problem, conducting both experimental studies and numerical simulations to unravel the underlying mechanisms. The experiments provide valuable insights into the heat transfer characteristics, flow patterns, and thermal performance, while numerical simulations offer a platform to explore the phenomena in a controlled and reproducible manner. Together, these studies have contributed to a deeper understanding of mixed convection heat transfer in the horizontal annulus with rotating inner cylinders, fostering advancements in diverse fields of application~\cite{fenot2011review}.

In the absence of cylinder rotation, heat transfer occurs solely due to natural convection in a concentric annulus. Extensive investigations of this geometry have been carried out using both experimental and numerical approaches. To name a few, for instance, Kuehn \& Goldstein~\cite{kuehn} conducted a detailed study with air and water, covering a wide range of Rayleigh numbers under steady flow conditions. They used the Mach-Zehnder interferometer technique for temperature evaluations and local heat transfer coefficient measurements, complemented by numerical investigations using the finite-difference method. Their results demonstrated excellent agreement between experimental and simulated temperature and velocity distributions at various positions within the annulus. Kumar~\cite{kumar1988study} conducted a numerical study encompassing a broad range of Rayleigh numbers, from conduction to convection-dominated steady flow states, while varying the diameter ratio between 1.2 and 10. Consistent with prior experiments, he observed crescent-shaped eddy patterns at small diameter ratios and kidney-shaped flow patterns at large diameter ratios. Yoo~\cite{yoo1999prandtl} performed numerical investigations to examine the impact of the Prandtl number on bifurcation phenomena and flow patterns within the annulus. This study revealed the existence of `upward flow' and `downward flow' depending on the value of the Prandtl number. Desai and Vafai~\cite{desai1994investigation} conducted a numerical investigation on turbulent natural convection inside a horizontal annulus across a wide range of Rayleigh and Prandtl numbers and radius ratios. They found a significant increase in heat transfer rate during the transition from laminar to turbulent flow, and this transition was delayed with increasing Prandtl numbers. Khanafer et al.~\cite{khanafer} studied the influence of a porous medium inside the annulus on natural convection heat transfer, while Nada and Said~\cite{nada2019effects} investigated the effects of fins and their configurations on the same. A comprehensive review of the studies for this particular problem can be found in the literature~\cite{powe,dawood2015forced}, thereby contributing to a better understanding of this mode of heat transfer in the concentric annulus geometry.

On the other hand, numerous investigations have also focused on forced convective heat transfer within this concentric annulus geometry with a rotating cylinder, employing a combination of experimental and simulation approaches. Aoki et al.~\cite{aoki} conducted both theoretical and experimental studies on convective heat transfer in this configuration, specifically involving a rotating and heated inner cylinder. They identified a significant rise in the Nusselt number beyond a critical value of the Taylor number. Their theoretical predictions aligned well with experimental observations, and they proposed an empirical correlation for the Nusselt number. Lee~\cite{lee1} conducted numerical simulations, maintaining fixed values for the Prandtl number of 0.7 and radius ratio of 2.6 while varying the Rayleigh numbers and rotational speeds. The heat transfer rate increased with the Rayleigh number at any rotational speed but decreased with the rotational speed at a fixed Rayleigh number. In a subsequent study~\cite{lee2}, Lee further explored the behavior at lower Prandtl numbers ranging from 0.01 to 0.1. At these low Prandtl numbers, the heat transfer rate from the inner rotating cylinder exhibited almost no dependence on the rotational speed or rotational Reynolds number. Gardiner \& Sabersky~\cite{gardiner} conducted experimental research, reaching high Taylor numbers of $10^6$ and rotational Reynolds numbers of up to 7000. They observed a sudden increase in the heat transfer coefficient. For a comprehensive review of forced convective heat transfer in concentric annulus configurations with a rotating cylinder, Childs et al.~\cite{childs1996review} have presented an extensive overview of this topic. 

The flow field and heat transfer phenomena in this concentric annulus geometry with a rotating cylinder become significantly more complex and intriguing when the buoyancy forces resulting from the temperature difference between the two cylinders are taken into account. Several studies have investigated this mixed convection heat transfer phenomenon in this particular system. For example, Fusegi et al.~\cite{fusegi} conducted a numerical study spanning a range of Grashof numbers and mixed convection parameters (the ratio of Grashof and square of Reynolds numbers) between $\infty$ (indicating pure natural convection) and 1. They found that increasing the mixed convection parameter led to higher heat transfer rates. Yoo~\cite{yoo} performed a numerical investigation on the concentric geometry with a cooled outer rotating cylinder, exploring a wide range of Rayleigh and Reynolds numbers for air with a Prandtl number of 0.7. They observed various flow patterns within the geometry, including one-eddy, two-eddies, and no-eddy configurations. Moreover, they noticed that the heat transfer rate decreased with the Reynolds number, regardless of the value of the Rayleigh number. Yang and Farouk~\cite{yang1992three} conducted a corresponding three-dimensional numerical study on mixed convection heat transfer in a concentric geometry with the inner rotating cylinder, varying the aspect ratios (the length to gap width ratio between outer and inner cylinders). They found that the heat transfer rate reached an asymptotic value with the aspect ratio, independent of the mixed convection parameter. Kahveci~\cite{kahveci2016stability} demonstrated that the flow field inside the annulus during mixed convection becomes unstable once the Rayleigh number surpasses a critical value. An increase in the Reynolds number and a decrease in the gap between the two cylinders facilitated this transition from stable to unstable flows.     

The aforementioned literature indicates a considerable number of studies on forced or mixed convection heat transfer in concentric annulus geometry with a rotating cylinder. However, most of these studies focused on simple Newtonian fluids like air or water. In contrast, there is a significant lack of research on non-Newtonian fluids, particularly viscoelastic fluids. Many fluids, such as polymer melts, solutions, suspensions, biofluids like blood, saliva, and synovial fluid, exhibit diverse non-Newtonian characteristics~\cite{chhabra}. These fluids find extensive applications in various industries, including food, pharmaceuticals, cosmetics, and polymer processing, where they display both viscous and elastic behaviors, known as viscoelastic behavior~\cite{morrison, thien}. Despite that there is almost no study available for viscoelastic fluids in the literature. This is probably because of the presence of several challenges associated with viscoelastic fluid simulations. One of the main challenges in studying viscoelastic fluids numerically is the High-Weissenberg Number Problem (HWNP), commonly encountered during simulations~\cite{keunings}. In this problem, the numerical solution diverges beyond a critical value of the Weissenberg number (denoted as $Wi=\lambda \dot{\gamma}$, where $\lambda$ is the fluid relaxation time, and $\dot{\gamma}$ is the strain rate) when dealing with flow through systems with geometric singularities. This divergence occurs due to the loss of positive definiteness in the stress tensor caused by steep gradients in variables like velocity and stress tensor. However, recent advancements, notably the log-conformation tensor approach introduced by Fattal \& Kupferman~\cite{fattal1}, have made it possible to simulate significantly large values of the Weissenberg number without encountering the divergence problem.

In the flow of viscoelastic fluids, elastic instability often arises when the Weissenberg number exceeds a critical value. This instability is triggered by streamline curvature and normal elastic stresses~\cite{larson, pakdel}. As the Weissenberg number increases further, the flow state transits to elastic turbulence (ET), a more chaotic and turbulent-like flow state~\cite{steinberg, groisman1, groisman2}. Research has shown that both elastic instability and elastic turbulence can significantly enhance heat transfer rates~\cite{whalley,li,traore,yao,poole} and mixing efficiency~\cite{burghelea2004mixing,groisman2001efficient,grilli2013transition}, especially in microfluidic systems where steady and laminar flow conditions prevail~\cite{sasmal2023applications}. In the context of mixed convection heat transfer, a recent study by Gupta et al.~\cite{gupta} revealed that the chaotic and fluctuating flow field resulting from elastic turbulence in viscoelastic fluids increased the heat transfer rate by more than 100\% compared to simple Newtonian fluids in a lid-driven cavity. Furthermore, their subsequent study~\cite{gupta2023effect} investigated how the aspect ratio of the lid-driven cavity influenced the flow dynamics and heat transfer rate due to mixed convection. These findings highlight the potential of elastic instability and elastic turbulence in enhancing transport processes, particularly in microfluidic systems where steady and laminar flow conditions predominantly exist.

Some further research has also explored the impact of fluid viscoelasticity on heat transfer phenomena. For example, Cheng et al.~\cite{cheng2017effect} investigated the influence of fluid viscoelasticity on Rayleigh-Bénard convection (RBC) within a square cavity. Their findings revealed a non-monotonic relationship between fluid viscoelasticity and heat transfer rates. Specifically, heat transfer rates initially decreased and then increased as the Weissenberg number increased. They conducted a turbulent kinetic energy (TKE) budget analysis to elucidate this trend. At low Weissenberg numbers, polymer additives acted as TKE sinks, leading to turbulent drag reduction (TDR) and heat transfer reduction (HTR). Conversely, at high Weissenberg numbers, polymer molecules acted as TKE sources, resulting in heat transfer enhancement (HTE). Other experimental~\cite{wei2012enhanced} and numerical~\cite{demir2003rayleigh} studies also observed these HTR and HTE phenomena, while some studies only reported HTR~\cite{cai2019polymer}. Consequently, a debate persists regarding whether fluid viscoelasticity induced by polymer additives diminishes or enhances heat transfer rates. This ongoing debate underscores the significance of investigating convective heat transfer in viscoelastic fluids, motivating the present study.

In particular, this study aims to conduct a numerical investigation encompassing forced, free, and mixed convection heat transfer in viscoelastic fluids confined between two concentric horizontal cylinders, with the inner cylinder in rotation. Our investigation not only focuses on analyzing flow dynamics and heat transfer but also includes a detailed analysis of the viscoelastic kinetic energy budget to enhance our understanding of flow and heat transfer physics. We will employ the finite volume method based open-source code OpenFOAM~\cite{openfoam} and RheoTool~\cite{rheotool} to solve the governing equations, including mass, momentum, and energy equations, while considering the Oldroyd-B viscoelastic constitutive equation to mimic the rheological behaviour of the present viscoelastic fluid. The rest of the paper is organized as follows: Section~\ref{sec:2} provides problem details and governing equations, Section~\ref{sec:3} outlines the numerical methodology, Section~\ref{sec:4} presents and discusses the results, and finally, Section~\ref{sec:5} summarizes the key findings of this study.

\section{\label{sec:2}Problem description and governing equations}

\begin{figure*}
    \centering
    \includegraphics[trim=0cm 0cm 0cm 0cm,clip,width=16cm]{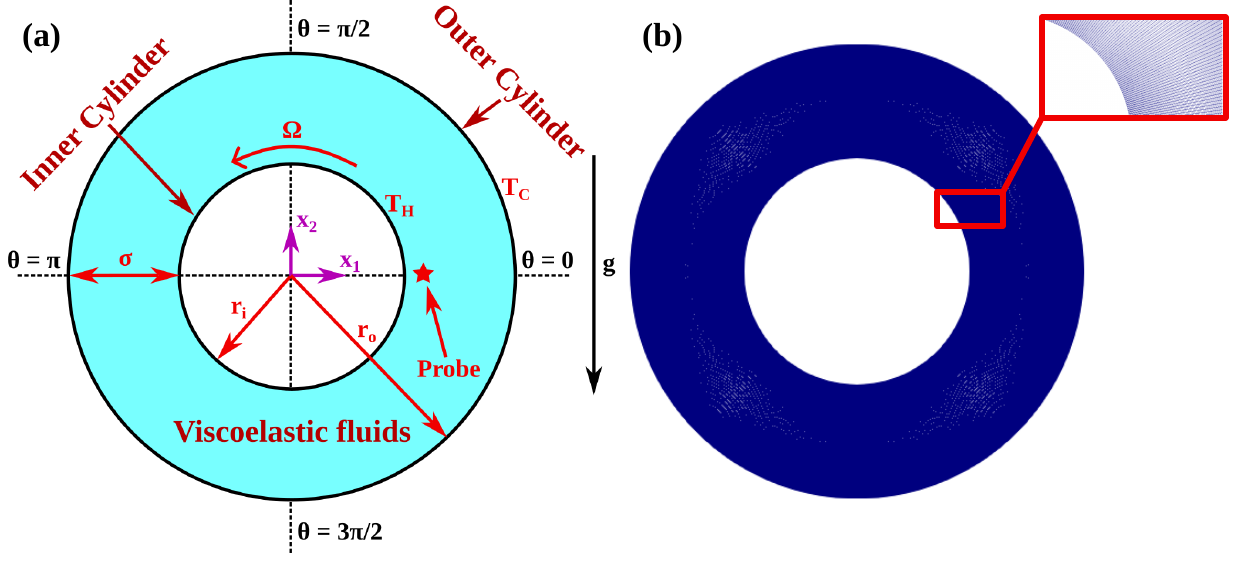}
    \caption{Schematic of the present problem (a), and the hexahedral grid used in the present study (b). Note that the probe within the annulus is located near the inner rotating cylinder at $(x_1,x_2)=(1.2,0)$.} 
    \label{fig:Geometry}
\end{figure*}

This study thoroughly explores thermal convection phenomena within a system consisting of two infinitely long horizontal concentric cylinders filled with viscoelastic fluids. The inner cylinder, with a radius of $r_{i}$, rotates at an angular velocity $\Omega$, while the outer cylinder, with a radius of $r_{o}$, remains fixed, as illustrated in Figure~\ref{fig:Geometry}(a). This configuration leads to a ratio of inner cylinder diameter to annular gap width denoted as $\sigma_{0} = \frac{2 r_{i}}{r_{o} - r_{i}}$. The inner cylinder is kept at a higher temperature $T_{H}$ compared to the temperature $T_{C}$ of the outer cylinder ($T_{H} > T_{C}$). This temperature difference generates buoyancy-induced convection within the viscoelastic fluid kept between the two cylinders, while the rotating inner cylinder induces rotation-induced convection.

The present study examines all three modes of thermal convection: forced convection (rotation-induced convection only), free convection (buoyancy-induced convection only), and mixed convection (both buoyancy-induced and rotation-induced convection). In the presence of buoyancy-induced thermal convection, the density variation is considered using the Boussinesq approximation: $\rho = \rho_{ref}[1-\beta_T(T-T_{ref})]$, where $\rho_{ref}$ is the reference density of the fluid at a reference temperature $T_{ref}$, and $\beta_T$ represents the thermal volumetric expansion coefficient at constant pressure, defined as $\beta_T = - \frac{1}{\rho} \frac{\partial \rho}{\partial T}|_{p}$. In this study, the reference temperature is set as the temperature of the outer cylinder, i.e., $T_{ref}=T_C$. To maintain the validity of the Boussinesq approximation, a small temperature difference ($< 5$ K) is applied between the inner and outer cylinder surfaces~\cite{bejan}.

Additionally, it is assumed that the thermophysical properties of the viscoelastic fluid, such as specific heat capacity $(C_p)$, thermal conductivity $(k)$, polymer relaxation time $(\lambda)$, zero-shear rate viscosity $(\eta_0)$, etc., remain independent of temperature. The Oldroyd-B viscoelastic constitutive equation is used to represent the rheological behavior of the fluid~\cite{bird}, chosen for several reasons: (i) simplicity with only two parameters - polymer concentration and relaxation time, (ii) development based on the simplest polymer kinetic theory assuming a dumbbell-like polymer molecule with two beads connected by an infinitely stretchable elastic spring, and (iii) ability to accurately replicate the rheological behavior of constant shear viscosity viscoelastic fluids (or the so-called Boger fluids~\cite{james}), commonly employed for studying the explicit influence of fluid elasticity on flow dynamics and heat transfer. Under these assumptions, the governing equations in their dimensional forms are expressed as follows:

Continuity equation:
\begin{equation} \label{eq:continuity}
   \frac{\partial {u}^{*}_i}{\partial x^{*}_i} = 0
\end{equation}

Momentum equation: 
\begin{equation} \label{eq:momentum}
    \rho \left ( \frac{\partial {u}^{*}_i}{\partial t^{*}} + {u}^{*}_j \frac{\partial {u}^{*}_i}{\partial x^{*}_j} \right) = - \frac{\partial p^{*}}{\partial x^{*}_i} + \eta_{s} \frac{\partial}{\partial x^{*}_j}\left (\frac{\partial {u}^{*}_i}{\partial x^{*}_j} \right)+\frac{\partial {\tau}^{p*}_{ij}}{\partial x^{*}_j} + \rho g \beta_T (T-T_{ref})\delta_{i2}
\end{equation}

Energy equation: 
\begin{equation} \label{eq:energy}
    \rho C_p\left ( \frac{\partial T}{\partial t^{*}} + {u}^{*}_j \frac{\partial T}{\partial x^{*}_j} \right) = k \frac{\partial}{\partial x^{*}_j}\left (\frac{\partial T}{\partial x^{*}_j} \right)
\end{equation}

In the above equations, ${u}^{*}_i$ is the velocity, $x^{*}_i$ is the position, $t^{*}$ is time, $p^{*}$ is the pressure, $\eta_s$ is the solvent viscosity, ${\tau}^{p*}_{ij}$ is the polymeric extra stress tensor, $g$ is the acceleration due to gravity, $T$ is the temperature, and $\delta_{ij}$ is the Kronecker delta. The polymeric extra stress tensor, ${\tau}^{p*}_{ij}$, for an Oldroyd-B viscoelastic fluid model is evaluated using the following equation~\cite{bird}:
\begin{equation} \label{eq:tau_p}
   {\tau}^{p*}_{ij} = \frac{\eta_p}{\lambda} ({C}_{ij}-{\delta}_{ij})
\end{equation}

\begin{equation} \label{eq:C_ij}
   {C}_{ij} + \lambda \overset{\nabla}{{C}_{ij}} = {\delta}_{ij}   
\end{equation}
Here $\eta_p$ is the polymeric viscosity, $\lambda$ is the polymer relaxation time, ${C}_{ij}$ is the polymeric conformation tensor, and $\overset{\nabla}{{C}_{ij}}$ is the upper-convected derivative of ${C}_{ij}$, which is given by the following equation:

\begin{equation} \label{eq:Conformation_transport}
       \overset{\nabla}{{C}_{ij}}= \dfrac{\partial{{C}_{ij}}}{\partial{t}^{*}}+{u}^{*}_{k}\frac{\partial {C}_{ij}}{\partial x^{*}_k}-\frac{\partial {u}^{*}_{i}}{\partial x^{*}_k}{C}_{kj}-\frac{\partial {u}^{*}_{j}}{\partial x^{*}_k}{C}_{ik}
\end{equation}
Next, we non-dimensionalize all these aforementioned governing  equations using the following scaling variables:

$x_i = \frac{x^{*}_i}{\sigma}$, $u_i = \frac{u^{*}_i}{\Omega r_i}$, $t = \frac{t^{*} \Omega r_i}{\sigma}$, $p = \frac{p^{*}}{\rho_{ref} {\Omega}^2 {r_i}^2}$, $\tau^{p}_{ij}=\frac{\tau^{p*}_{ij} \sigma}{\eta_p \Omega r_i}$ and $\phi = \frac{T-T_C}{T_H-T_C}$. 
\\ 
The non-dimensional forms of the above governing equations are written as below:

Continuity equation:
\begin{equation}
   \frac{\partial {u_i}}{\partial x_i} = 0
\end{equation}

Momentum equation: 
\begin{equation}
     \frac{\partial {u_i}}{\partial t} + {u_j} \frac{\partial {u_i}}{\partial x_j} = - \frac{\partial p}{\partial x_i} + \frac{\beta}{Re} \frac{\partial}{\partial x_j}\left (\frac{\partial {u_i}}{\partial x_j} \right)+\frac{1-{\beta}}{Re}\frac{\partial {\tau^{p}_{ij}}}{\partial x_j} + \frac{Gr}{Re^2} \phi \delta_{i2}
\end{equation}

Energy equation: 
\begin{equation}
     \frac{\partial \phi}{\partial t} + {u_j} \frac{\partial \phi}{\partial x_j} = \frac{1}{Re Pr}\frac{\partial}{\partial x_j}\left (\frac{\partial \phi}{\partial x_j} \right)
\end{equation}

Polymeric conformation tensor transport equation:
\begin{equation}
       \dfrac{\partial{{C}_{ij}}}{\partial{t}}+{u}_{k}\frac{\partial {C}_{ij}}{\partial x_k}-\frac{\partial {u}_{i}}{\partial x_k}{C}_{kj}-\frac{\partial {u}_{j}}{\partial x_k}{C}_{ik}=\frac{\delta_{ij}-{C}_{ij}}{Wi}
\end{equation}
In the above equations, $Re = \frac{\rho \Omega r_i \sigma}{\eta_0}$ is the Reynolds number (ratio of the inertial to that of the viscous forces), $Pr = \frac{C_p \eta_0}{k}$ is the Prandtl number (ratio of the momentum to that of the thermal diffusivity), $Gr = \frac{{\rho}^2 g \beta_T (T_H-T_C) {\sigma}^3}{{\eta_0}^2}$ is the Grashof number (ratio of the buoyancy to that of the viscous forces), $Wi = \frac{\lambda \Omega r_i}{\sigma}$ is the Weissenberg number (ratio of the elastic to that of the viscous forces), and $\beta = \frac{\eta_s}{\eta_0} = \frac{\eta_s}{\eta_s + \eta_p}$ is the polymer viscosity ratio (ratio of the solvent to that of the zero-shear rate viscosity of the polymer solution). As $\beta$ approaches 1, the fluid demonstrates Newtonian behavior; conversely, it takes on the characteristics of a polymer melt as $\beta$ approaches 0. In this study, we also employ another dimensionless parameter known as the Richardson number $(Ri)$, defined as $Ri = \frac{Gr}{Re^2}$. In general, when $Ri < 0.1$, heat transfer is predominantly driven by forced convection. Conversely, for $Ri > 10$, free convection dominates significantly. Mixed convection heat transfer occurs between these two limits, with a Richardson number ranging from $0.1$ to $10$. The above set of equations is valid for purely forced and mixed convection modes of heat transfer. In the limit of $Ri \rightarrow 0$, the equations will reduce to that of purely forced convection heat transfer. However, in the case of purely free convection, a different set of equations will govern the flow field and heat transfer, presented in Appendix~\ref{App}. This is because the scaling variable for the velocity is different in this mode of heat transfer compared to that of forced and mixed convection due to the absence of cylinder rotation.

\section{\label{sec:3}Numerical methodology}

\subsection{Simulation procedure}

In this study, we have employed the finite volume method (FVM) based open-source computational fluid dynamics (CFD) code OpenFOAM (version 7)~\cite{openfoam} to solve all the governing equations, namely, mass, momentum, and energy equations. Additionally, we have utilized the recently developed~\textit{rheoHeatFoam} solver available in the RheoTool (version 5) package~\cite{rheotool} to solve the Oldroyd-B viscoelastic constitutive equation. Among various discretization options, we have opted for the high-resolution CUBISTA scheme for discretizing advective terms in momentum, energy, and constitutive equations due to its improved iterative convergence properties~\cite{cubista}. We have utilized the second-order accurate Gauss linear orthogonal interpolation scheme for diffusion terms in both momentum and energy equations. Gradient terms were discretized using the Gauss linear corrected scheme, and the Euler time integration scheme was applied for time derivative terms. To solve the linear system of pressure fields, we have employed the preconditioned conjugate solver (PCG) with a DIC (diagonal-based incomplete Cholesky) preconditioner~\cite{DIC}. For solving velocity, temperature, and stress fields, the preconditioned bi-conjugate gradient solver (PBiCG) with a DILU (diagonal-based incomplete LU) preconditioner was utilized~\cite{DILU}. The pressure-velocity coupling was accomplished using the SIMPLE method. To enhance numerical stability, we have implemented the log-conformation tensor approach, initially introduced by Fattal and Kupferman~\cite{fattal2} and later integrated into OpenFOAM by Pimenta and Alves~\cite{pimenta}. Furthermore, we set a relative tolerance level of $10^{-8}$ for velocity, pressure, temperature, and stress fields.

\subsection{Boundary and initial conditions}
We have employed the following set of boundary and initial conditions for the numerical solution of the current problem:

\textbf{Inner cylinder surface:} An anticlockwise angular velocity of $\Omega$ is imposed. The temperature is kept constant at $\phi = 1$. The polymeric extra stresses are extrapolated linearly to the surface, and a zero-pressure gradient is applied.

\textbf{Outer cylinder surface:} The standard no-slip and no-penetration $(u_i = 0)$ conditions for velocity are applied. Additionally, linear extrapolation is employed for polymeric extra stresses, and a zero-pressure gradient is maintained while the temperature is set to a fixed value of $\phi = 0$.

Furthermore, initial conditions involve all variables, including velocity vector, temperature, and polymeric extra stress tensor, being set to zero.

\subsection{Grid and time independence studies}
\begin{table}
\caption{\label{table:Grid}Details of the grid independence study performed at $Re = 1000$, $Ra = Gr \times Pr = 10^{6}$, and $\sigma_0 = 2$. Here, $N_{\theta}$ and $N_{r}$ denote the number of elements on the surface of the inner cylinder and in the radial direction, respectively. The $<.>$ operator represents the time-averaged quantity.}
\begin{ruledtabular}
\setlength{\extrarowheight}{2pt}
\begin{tabular}{cccccccc}
\multirow{2}{*}{Grid Type} & \multirow{2}{*}{$N_{\theta}$} & \multirow{2}{*}{$N_{r}$} & \multirow{2}{*}{Total elements} & \multicolumn{2}{c}{$\textit{Wi}=5$} & \multicolumn{2}{c}{$\textit{Wi}=10$}\\
 &  & & & $<Nu_{avg}>$ & \% error & $<Nu_{avg}>$ & \% error\\[0.1cm] 
\hline
G1 & 300 & 100 & 30,000 & 1.607 & -- & 2.445 & --\\
G2 & 432 & 140 & 60,400 & 1.623 & 0.996 & 2.644 & 8.139\\
G3 & 600 & 200 & 120,000 & 1.635 & 0.739 & 2.604 & 1.513\\
G4 & 800 & 300 & 240,000 & -- & -- & 2.609 & 0.192\\
\end{tabular}
\end{ruledtabular}
\end{table}
\begin{figure}
    \centering
    \includegraphics[trim=0cm 0cm 0cm 0cm,clip,width=16cm]{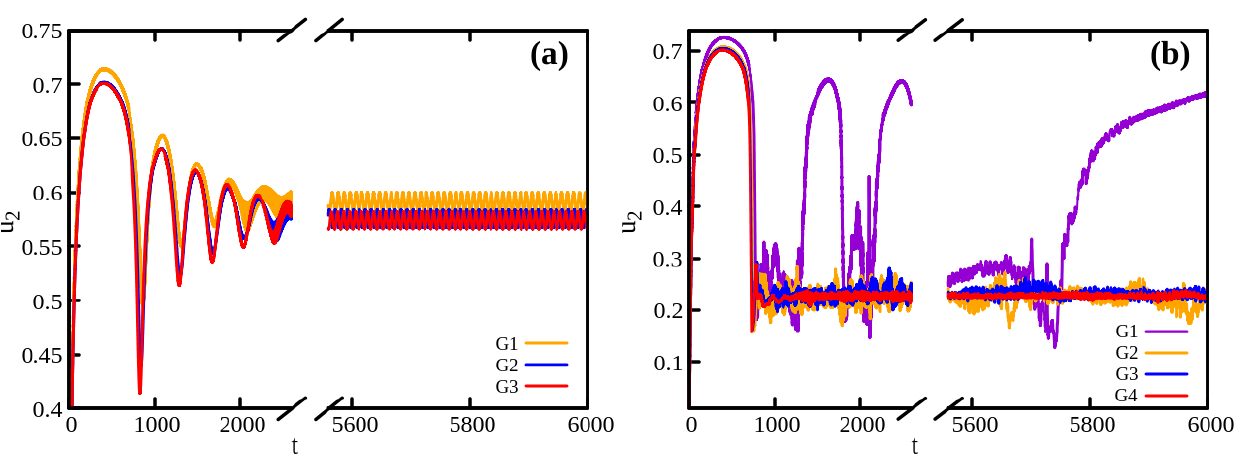}
    \caption{Temporal variation of the non-dimensional velocity component at a probe location near the inner cylinder surface $(x_1=1.2,x_2=0)$ for different grids considered in this study at $Wi=5$ (a), and $Wi=10$ (b).}
    \label{fig:GridTest}
\end{figure}

In addition to meticulous choices of discretization techniques, tolerance thresholds, and boundary conditions, it is imperative to select an optimal grid density that upholds result accuracy and precision in any computational fluid dynamics analysis. The time step size is equally influential in problem resolution. In this study, we have followed the established procedure of employing the `blockMeshDict' subroutine in OpenFOAM to construct the computational domain and its meshing using regular hexahedral cells. Ensuring stability and accuracy, we have maintained the Courant number ($Co = \frac{u \Delta t}{\Delta x}$) below unity to satisfy the Courant-Friedrichs-Lewy (CFL) condition. Here, $\Delta t$ represents the time step size, and $\frac{\Delta x}{u}$ represents the characteristic convective time scale. We have observed $Co \le 0.3$ in all simulations when employing $\Delta t \le 10^{-3}$. After identifying an appropriate time step size for capturing temporal variations in fields like velocity, temperature, and stress, we have conducted a grid independence study. It is well established that as Reynolds and/or Grashof numbers increase, the hydrodynamic and/or thermal boundary layer thickness decreases, leading to steeper gradients near the inner and outer cylinder walls. To address this, grid refinement near these surfaces (as illustrated in Fig.~\ref{fig:Geometry}(b)) becomes essential. Consequently, we have conducted a grid independence study at maximum Reynolds, Grashof, and Weissenberg numbers to capture steep gradients in thin boundary layers. For this study, we have considered Weissenberg numbers 5 and 10, presenting the corresponding time-averaged inner cylinder surface Nusselt numbers in Table~\ref{table:Grid}. After assessing the time-averaged Nusselt numbers and temporal velocity variation at a probe location (refer to Fig.~\ref{fig:GridTest}), we have selected grid G2 with 60,480 hexahedral cells for investigations up to $Wi=5$. Additionally, grid G3 with 120,000 hexahedral cells proved sufficient for Weissenberg numbers ranging from 5 to 10. We have adopted two grids due to the transition to turbulent-like behavior as Weissenberg numbers exceeded 5. Increasing grid resolution in such conditions exacerbated fluctuations in velocity, temperature, stress, etc., making grid independence unattainable. An attempt to minimize error involved creating grid G4 with 240,000 cells, confirming convergence and reliability (within 2\% of difference) of power spectral density (PSD) slopes for velocity and temperature fluctuations across grids G3 to G4 (results are not shown here). Notably, a Hopf bifurcation precedes the transition to elastic turbulence, with this phenomenon emerging at approximately $Wi \approx 3$ in our study. As $Wi$ reaches 5, the periodicity amplitude increases, as evident in sub-Fig.~\ref{fig:GridTest}(a). Comparing the patterns and velocity component values between grids G2 and G3, we have validated G2's suitability for $Wi\le 5$. In summary, our results rest on grid G2 for $Wi \le 5$ and grid G3 for $5 < Wi \le 10$, with a time step size of $\Delta t \le 10^{-3}$.

\section{\label{sec:4}Results and discussions}
In this study, we have conducted comprehensive numerical simulations to investigate thermal convection phenomena in viscoelastic fluids confined between two concentric horizontal cylinders. The parameters investigated included Reynolds number ($Re = 1000$), Grashof number ($0 \le Gr \le 1.428 \times 10^5$), Weissenberg number ($0 \le Wi \le 10$), polymer viscosity ratio ($\beta = 0.5$), Prandtl number ($Pr = 7$), and a fixed value of the ratio between inner cylinder diameter to annulus gap width ($\sigma_0 = 2$). The simulations covered forced, free, and mixed modes of heat transfer, facilitating a comprehensive comparative analysis. The rheological behavior of the viscoelastic fluid was described using the Oldroyd-B constitutive equation, while the log-conformation tensor approach was employed to handle challenges related to the High Weissenberg Number Problem (HWNP) that often arises in viscoelastic fluid simulations. Additionally, simulations were carried out for a Newtonian fluid ($Wi = 0$ and $\beta = 1$) under identical conditions to discern the impact of fluid viscoelasticity on flow dynamics and heat transfer. The obtained results were discussed in terms of velocity component variations, velocity magnitude, streamlines, isotherms, averaged Nusselt numbers, and viscoelastic kinetic energy budget analysis. Before presenting and discussing these new findings, we have ensured the accuracy and reliability of our present numerical tool by conducting validation studies against existing results that are available in the literature.

\subsection{Code validation}

\begin{figure}
    \centering
    \includegraphics[trim=0cm 0cm 0cm 0cm,clip,width=16cm]{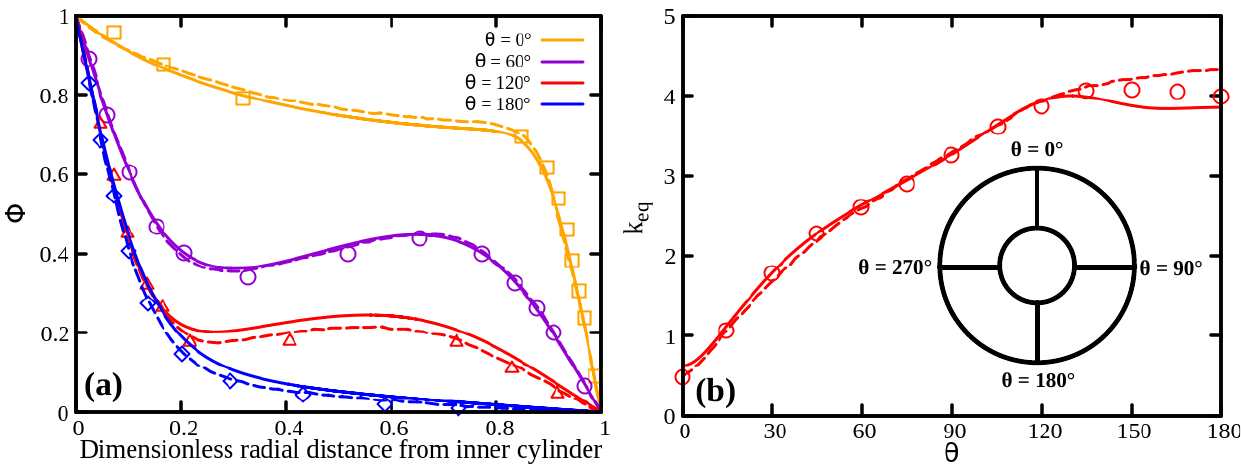}
    \caption{Comparison of the non-dimensional temperature variation along the dimensionless radial distance from the inner cylinder (a), and the local equivalent thermal conductivity of the inner cylinder (b) between the present results (solid lines) with that of experimental (symbols) and numerical results (dashed lines) of Kuehn \& Goldstein~\cite{kuehn} for free convection case. The parameters used are $Ra = 5 \times 10^4$, $Pr = 0.7$, and $\sigma_{0} = 1.25$.} 
    \label{fig:validation_paper3}
\end{figure}

\begin{figure}
    \centering
    \includegraphics[trim=0cm 0cm 0cm 0cm,clip,width=16cm]{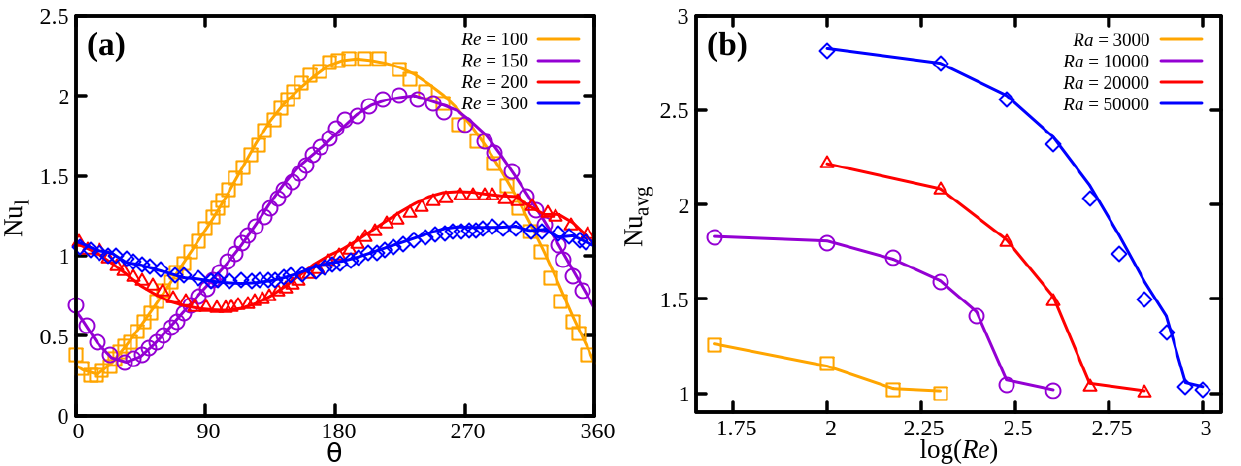}
    \caption{Comparison of the local Nusselt number along the surface of the inner cylinder at $Ra=5000$ (a), and average Nusselt number (b) between the present results (solid lines) with that of numerical results (symbols) of Yoo~\cite{yoo} for $Pr = 0.7$ and $\sigma_{0} = 2$. In this case, the azimuthal coordinate $\theta$ was measured counter-clockwise from the upward vertical through the center of the cylinders.} 
    \label{fig:validation_paper1}
\end{figure}

First, we have verified the accuracy of our numerical code by comparing the results obtained with it against experimental and numerical results provided by Kuehn and Goldstein~\cite{kuehn} for the pure free convection case between two horizontal concentric cylinders using air as the working fluid. The comparison, depicted in sub-Fig.~\ref{fig:validation_paper3}(a), illustrates the non-dimensional temperature distribution variations along the dimensionless radial distance at various circumferential positions. Additionally, the local equivalent thermal conductivity of the inner cylinder along the circumferential direction is shown in sub-Fig.~\ref{fig:validation_paper3}(b). The agreement between our results and those of Kuehn and Goldstein~\cite{kuehn} is acceptably close. Subsequently, we have validated our numerical tool for the mixed convection case within the same configuration of two horizontal concentric cylinders, using numerical outcomes provided by Yoo et al.~\cite{yoo}. In sub-Figs.~\ref{fig:validation_paper1}(a) and (b), we have presented comparisons of the variation in local and average Nusselt numbers at different Reynolds numbers, maintaining a constant Rayleigh number of 5000. In both instances, a good concurrence is evident between our results and those presented by Yoo et al.~\cite{yoo}.

\subsection{Flow dynamics}

\begin{figure}
        \centering
        \includegraphics[trim=0cm 0cm 0cm 0cm,clip,width=13.75cm]{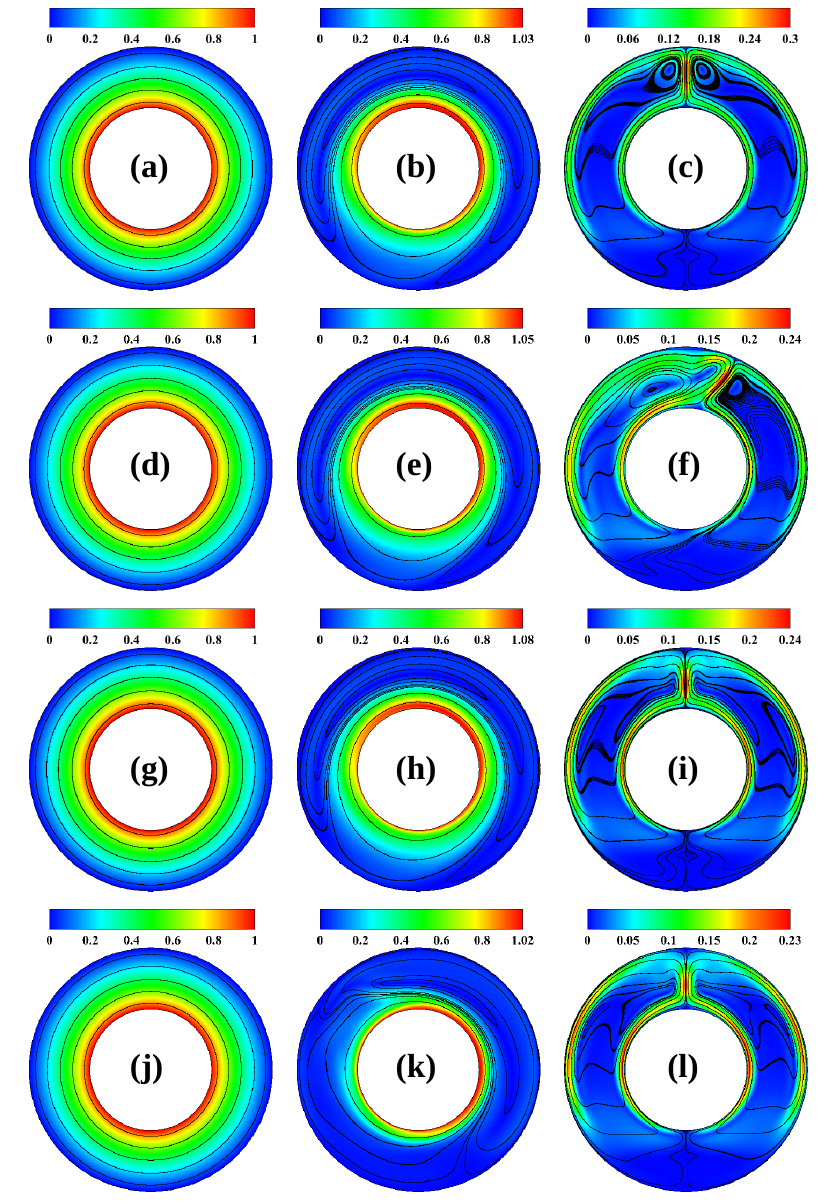}
        \caption{Time-averaged non-dimensional velocity magnitude contours and streamline patterns for different types of convection phenomena at $\textit{Wi}=0$, $\textit{Ri}=0$ (a); $\textit{Wi}=0$, $\textit{Ri}=0.143$ (b); $\textit{Wi}=0$, $\textit{Ri}\rightarrow \infty$ (c); $\textit{Wi}=1$, $\textit{Ri}=0$ (d); $\textit{Wi}=1$, $\textit{Ri}=0.143$ (e); $\textit{Wi}=1$, $\textit{Ri}\rightarrow \infty$ (f); $\textit{Wi}=5$, $\textit{Ri}=0$ (g); $\textit{Wi}=5$, $\textit{Ri}=0.143$ (h); $\textit{Wi}=5$, $\textit{Ri}\rightarrow \infty$ (i); $\textit{Wi}=10$, $\textit{Ri}=0$ (j); $\textit{Wi}=10$, $\textit{Ri}=0.143$ (k); and $\textit{Wi}=10$, $\textit{Ri}\rightarrow \infty$ (l).} 
        \label{fig:Streamlines}
    \end{figure}
    
At the onset, we present an analysis of flow dynamics aspects within the current system encompassing all three modes of thermal convection: forced, mixed, and natural convection. Figure~\ref{fig:Streamlines} depicts the non-dimensional velocity magnitude and streamline patterns for different values of the Richardson $(Ri)$ and Weissenberg $(Wi)$ numbers. Notably, forced convection occurs when $Ri = 0$ (first column in Fig.~\ref{fig:Streamlines}), while free convection becomes predominant as $Ri$ approaches infinity (last column in Fig.~\ref{fig:Streamlines}). Meanwhile, mixed convection is characterized by an intermediate value of the Richardson number (middle column in Fig.~\ref{fig:Streamlines}). Moreover, the case with $Wi = 0$ corresponds to that of a Newtonian fluid (first row in Fig.~\ref{fig:Streamlines}).

In the case of pure forced convection, the streamlines follow the body contours of the cylinders, and as a result, they appear to be concentric circles within the region between the two cylinders irrespective of the value of the Weissenberg number. It particularly shows the existence of a solid body type rotation inside the system, where the relative distance between any two fluid parcels will remain the same all the time. The fluid velocity is solely in the circumferential direction, and there is no velocity in the radial direction. Therefore, one can expect no convective mixing of hot and cold fluids present near the hot inner and cold outer cylinders. The heat transfer will predominantly occur by the conduction mode, which will be discussed in detail in the subsequent section. 

On the other hand, in the case of pure free convection, the fluid that is present at the bottom near the inner hot cylinder $(\theta = 3\pi/2)$ is heated up and becomes lighter so that it rises towards the top $(\theta = \pi/2)$. Therefore, it generates a buoyant plume that travels from the bottom region of the inner cylinder to the top region and forms a high-velocity magnitude zone in the middle of the upper gap region between the two cylinders. Ultimately, it touches the top region of the cold outer cylinder and travels down along the perimeter of the cylinder. Therefore, a buoyancy-induced circulation of hot and cold fluids is created inside the system. This leads to the mixing between them, and therefore, one can expect a greater heat transfer rate than pure forced convection. Furthermore, due to this upward movement of the buoyant plume near the hot cylinder and downward movement near the outer cylinder, the streamlines are found to be highly distorted in this case as compared to circular ones seen in the pure forced convection case. In particular, two recirculation regions are formed at the top of the gap between the two cylinders. Moreover, the flow remains steady and symmetric along the vertical line passing through the origin of the present flow system for Newtonian fluids (sub-Fig.~\ref{fig:Streamlines}(c)). 

However, the flow transits to an unsteady and asymmetric state for viscoelastic fluids when the Weissenberg number ($Wi$) is set to 1, as depicted in sub-Fig.~\ref{fig:Streamlines}(f). The region of high-velocity magnitude initially observed at $\theta = \pi/2$ for Newtonian fluids is now displaced towards the right-hand side of the origin for viscoelastic fluids. It is important to note that these streamlines and velocity magnitudes are determined based on time-averaged velocity fields. This asymmetry in the flow structure is apparent not only in the profiles of streamlines observed between the two halves of the system, but also in the distribution of velocity magnitudes. For instance, the velocity magnitude appears higher on the left-hand side of the flow system, particularly in the proximity of the inner and outer cylinders. As the Weissenberg number increases further to 5 (sub-Fig.\ref{fig:Streamlines}(i)), the flow again somehow transits back to a more symmetric state. The recirculation regions visible at the position $\theta = \pi/2$ for Newtonian fluids are not observed at this Weissenberg number for viscoelastic fluids. Additionally, the velocity magnitude near the inner and outer cylinders increases with the Weissenberg number, while it decreases in the upper gap region between the two cylinders. This trend becomes more pronounced as the Weissenberg number is further incremented to 10, as shown in sub-Fig.\ref{fig:Streamlines}(l).   

\begin{figure}
        \centering
        \includegraphics[trim=0cm 0cm 0cm 0cm,clip,width=12.5cm]{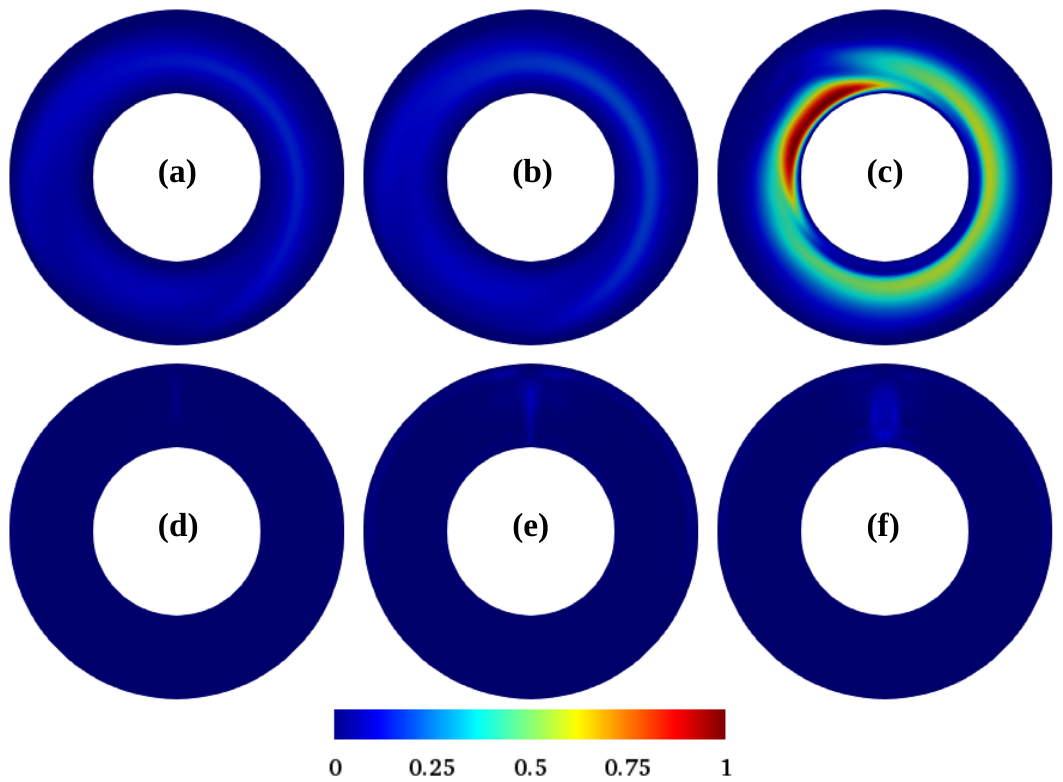}
        \caption{Variation of the root mean square velocity magnitude ﬂuctuations for different values of the Weissenberg number and Richardson number, namely, $\textit{Wi}=0$, $\textit{Ri}=0.143$ (a); $\textit{Wi}=5$, $\textit{Ri}=0.143$ (b); $\textit{Wi}=10$, $\textit{Ri}=0.143$ (c); $\textit{Wi}=0$, $\textit{Ri}\rightarrow \infty$ (d); $\textit{Wi}=5$, $\textit{Ri}\rightarrow \infty$ (e); and $\textit{Wi}=10$, $\textit{Ri}\rightarrow \infty$ (f). Here, the root mean square velocity magnitude is presented on a scale of 0 to 1 after rescaling with its maximum value obtained at $\textit{Wi}=10$, $\textit{Ri}=0.143$.} 
        \label{fig:UP2M}
\end{figure}

\begin{figure}
        \centering
        \includegraphics[trim=0cm 0cm 0cm 0cm,clip,width=16cm]{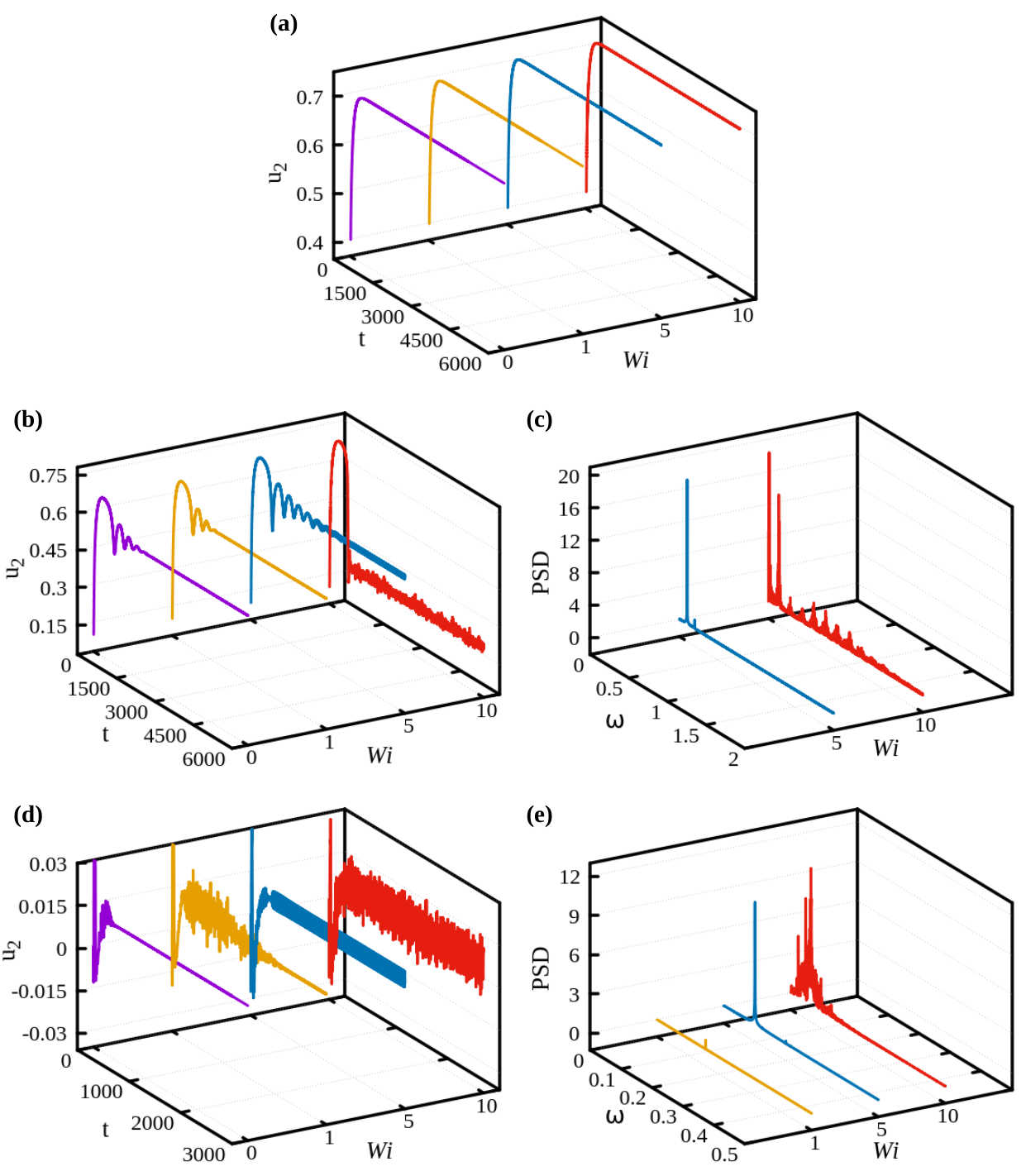}
        \caption{Temporal variation of the non-dimensional velocity component at a probe location near the inner cylinder surface $(x_1=1.2,x_2=0)$ for $\textit{Ri}=0$ (a), $\textit{Ri}=0.143$ (b), and $\textit{Ri}\rightarrow \infty$ (d). The corresponding power spectral density curves represent $\textit{Ri}=0.143$ (c), and $\textit{Ri}\rightarrow \infty$ (e).} 
        \label{fig:FFT}
\end{figure}

In the case of mixed convection, the flow dynamics within the system is governed by the interplay of buoyancy-driven and rotation-induced convection. Buoyancy-induced convection causes the fluid to move radially upward towards the upper gap region of the system, while rotation-induced convection pulls the fluid circumferentially. Consequently, these two convection mechanisms interact, with the strongest interaction occurring at the position $\theta = \pi/2$ near the inner cylinder. This interaction gives rise to a region of high-velocity magnitude for Newtonian fluids, as evidenced in sub-Fig.~\ref{fig:Streamlines}(b). Moreover, the streamlines resemble concentric patterns near the inner cylinder, reminiscent of the purely forced convection case due to the dominance of rotation-induced convection in this area. However, moving from the inner cylinder towards the outer cylinder leads to the loss of concentric streamline patterns, resulting in significant distortion. Particularly noteworthy is the formation of a curved kidney-shaped region in the upper gap between the two cylinders. With an increase in the Weissenberg number, the zone of high-velocity magnitude gradually shifts towards the $\theta = 0$ position. This trend is illustrated in the results at $Wi = 10$ in sub-Fig.~\ref{fig:Streamlines}(k). As the Weissenberg number increases, the hydrodynamic boundary layer thickness decreases, especially evident at the highest Weissenberg number considered in this study. Additionally, the streamlines exhibit increased distortion, and the kidney-shaped region's size decreases with the Weissenberg number in the mixed convection case. Hence, a significant difference in flow dynamics emerges within the present system depending upon the type of heat transfer mode and the Weissenberg number. This difference, in turn, substantially influences the ensuing heat transfer characteristics, an aspect that will be discussed in the subsequent section.

Therefore, in free and mixed convection conditions, the flow field tends to become unsteady and fluctuating in nature as the Weissenberg number gradually increases. This fluctuation in the flow field is demonstrated in Fig.~\ref{fig:UP2M} wherein the time-averaged root mean square velocity magnitude fluctuation is presented both for mixed (top row) and free (bottom row) convection cases. It is defined as $\left({u}_{rms,mag} = \sqrt{<(\Tilde{u}_{mag} - \Bar{u}_{mag})^{2}>_{t}} \right)$ where $\Tilde{u}_{mag}$ is the instantaneous velocity magnitude whereas $\Bar{u}_{mag}$ is its time-averaged value. It can be clearly seen that the fluctuation in the flow field increases with the Weissenberg number. Furthermore, it is less in the case of free convection than seen for the mixed convection case, particularly see the results presented in sub-Fig.~\ref{fig:UP2M}(c) at $Wi = 10$ for which a highly fluctuating flow field is observed.    

To characterize the nature of this fluctuating flow field under various heat transfer modes and different Weissenberg number values, Fig.~\ref{fig:FFT} illustrates the temporal evolution of the non-dimensional $u_{2}$ velocity component at a specific probe location near the inner cylinder (refer to sub-Fig.~\ref{fig:Geometry}(a)) and its corresponding power spectral density (PSD) plot. In the case of forced convection ($Ri = 0$), as depicted in sub-Fig.~\ref{fig:FFT}(a), the velocity component remains constant over time irrespective of the Weissenberg number. Consequently, the flow field remains steady for both Newtonian and viscoelastic fluids under forced convection. In contrast, in the case of mixed convection, the velocity component also maintains a constant value up to a Weissenberg number less than 3, indicating the presence of a steady flow field. However, at $Wi = 5$, the velocity component exhibits regular quasi-periodic fluctuations with time, and at $Wi = 10$, it displays more chaotic aperiodic fluctuations, as illustrated in sub-Fig.\ref{fig:FFT}(b). This behavior is further confirmed in sub-Fig.\ref{fig:FFT}(c), which presents the corresponding power spectral density plot of velocity fluctuations. At $Wi = 5$, the velocity fluctuations are characterized by a dominant primary frequency with a high amplitude and a secondary frequency with a relatively smaller amplitude, indicating a quasi-periodic flow state at this Weissenberg number. As the Weissenberg number increases to 10, many secondary frequencies emerge across a wide range of values alongside two dominant frequencies. This observation indicates an aperiodic and chaotic flow state at $Wi = 10$ and suggests the presence of elastic turbulence phenomenon within the system at this condition~\cite{steinberg,groisman1,groisman2}. In the case of pure free convection, the flow field exhibits unsteadiness even at lower Weissenberg number values compared to mixed convection (sub-Fig.~\ref{fig:FFT}(d)). Under this heat transfer mode, the flow field transits to an unsteady periodic state at $Wi = 1$, highlighted by the presence of a single dominant frequency peak in the PSD plot presented in sub-Fig.~\ref{fig:FFT}(e). As the Weissenberg number increases to 5, the flow field gradually shifts to a quasi-periodic state and eventually to an aperiodic state at $Wi = 10$, as evident in sub-Fig.~\ref{fig:FFT}(e). 

\begin{figure}
        \centering
        \includegraphics[trim=0cm 0cm 0cm 0cm,clip,width=14cm]{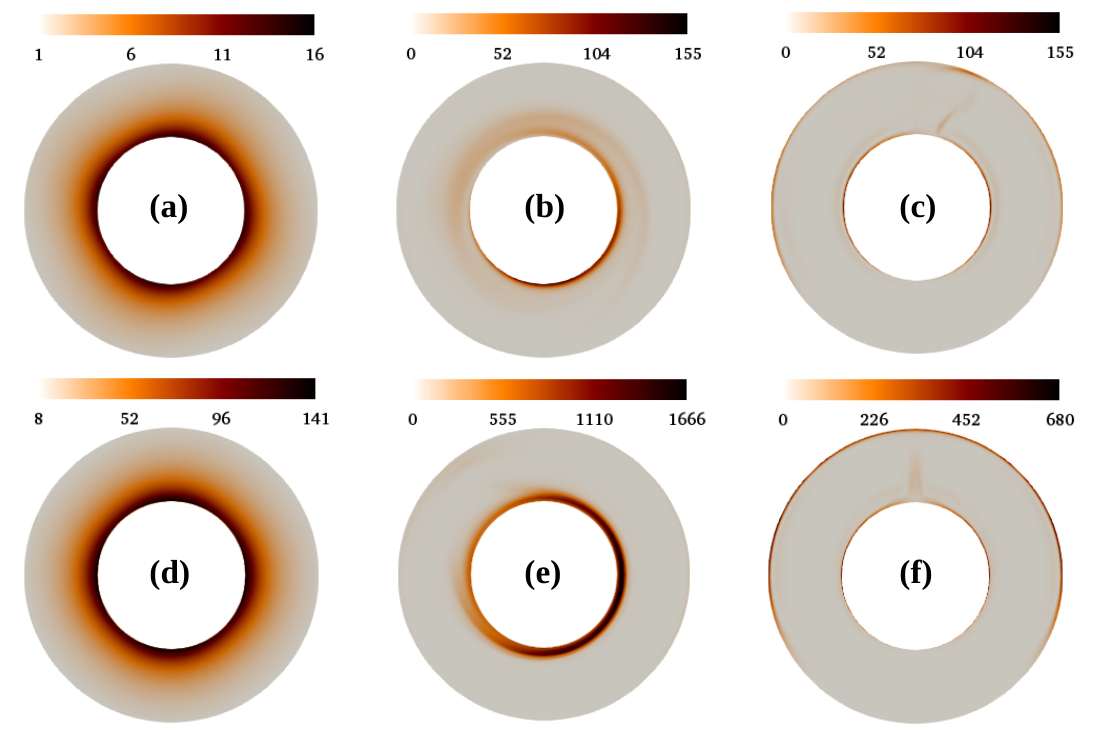}
        \caption{Variation of the time-averaged magnitude of  polymeric extra stress tensor $(|\tau_{ij}^{p}|)$ for different values of Weissenberg and Richardson numbers, namely, $\textit{Wi}=1$, $\textit{Ri}=0$ (a); $\textit{Wi}=1$, $\textit{Ri}=0.143$ (b); $\textit{Wi}=1$, $\textit{Ri}\rightarrow \infty$ (c); $\textit{Wi}=10$, $\textit{Ri}=0$ (d); $\textit{Wi}=10$, $\textit{Ri}=0.143$ (e); and $\textit{Wi}=10$ $\textit{Ri}\rightarrow \infty$ (f).} 
        \label{fig:Tau}
    \end{figure}

Therefore, for free and mixed convection heat transfer modes, a gradual transition in the flow field happens from steady to unsteady periodic to unsteady quasi-periodic and finally to a more chaotic aperiodic state as the Weissenberg number gradually increases in the case of viscoelastic fluids in comparison to a steady flow state for the Newtonian fluids. This transition occurs due to the occurrence of elastic instability in flows of viscoelastic fluids, which, as mentioned earlier, originates due to the interaction between the streamline curvature and normal elastic stresses present in a viscoelastic fluid~\cite{pakdel}. The magnitude of elastic polymeric stresses increases with the Weissenberg number, as shown in Fig.~\ref{fig:Tau}. It can be seen that the magnitude of polymeric stress is mostly higher in the vicinity of the inner cylinder, where severe stretching of polymer molecules happens due to its rotation. However, in the case of free convection, a high-magnitude stress zone is also observed near the outer cylinder due to the downward motion of the fluid in this region. Furthermore, the polymeric stress is much higher for the mixed convection case than for the cases of forced and free convection. Therefore, one would expect a much higher intensity in the elastic instability and subsequent elastic turbulence phenomena in the mixed convection case of heat transfer of viscoelastic fluids, which was also evident in the analysis of temporal velocity fluctuation at a probe location presented in Fig.~\ref{fig:FFT} and variation in the time-averaged root mean square velocity magnitude fluctuation shown in Fig.~\ref{fig:UP2M}.

\subsection{Heat transfer}

\begin{figure}[hbt!]
    \centering
    \includegraphics[trim=0cm 0cm 0cm 0cm,clip,width=13cm]{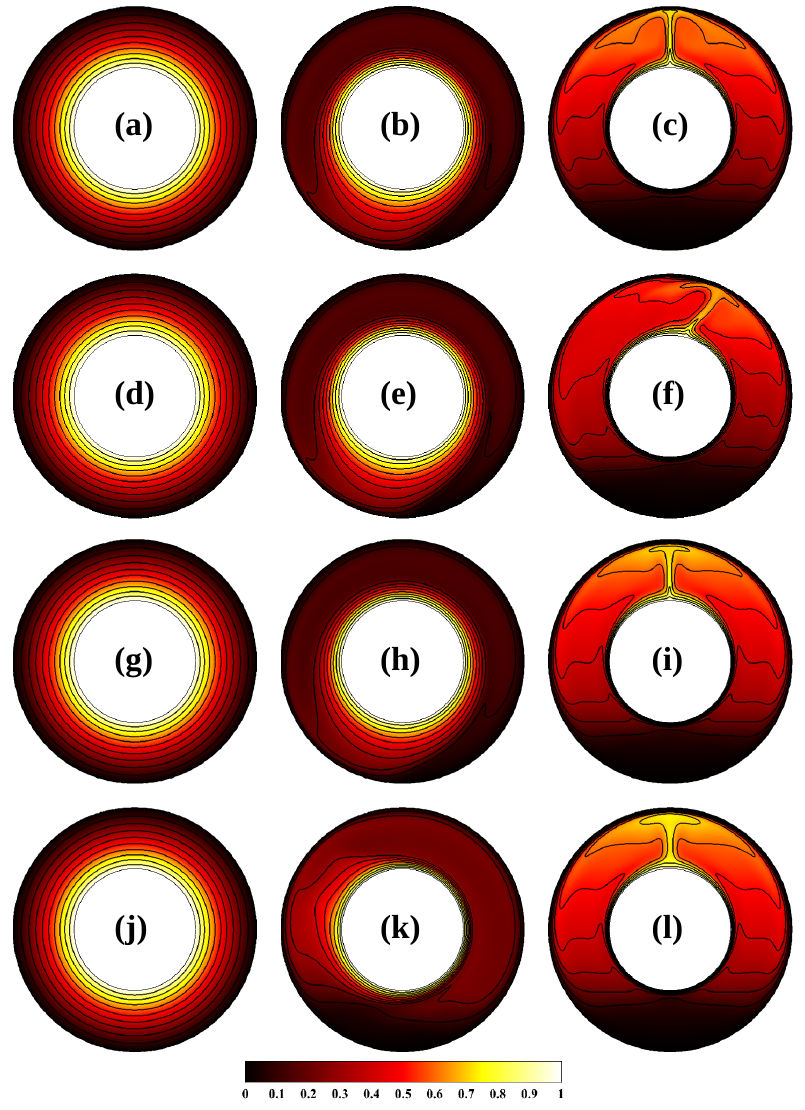}
    \caption{Time-averaged surface plot of temperature distribution and isotherm contours for various types of convection phenomena at $\textit{Wi}=0$, $\textit{Ri}=0$ (a); $\textit{Wi}=0$, $\textit{Ri}=0.143$ (b); $\textit{Wi}=0$, $\textit{Ri}\rightarrow \infty$ (c); $\textit{Wi}=1$, $\textit{Ri}=0$ (d); $\textit{Wi}=1$, $\textit{Ri}=0.143$ (e); $\textit{Wi}=1$, $\textit{Ri}\rightarrow \infty$ (f); $\textit{Wi}=5$, $\textit{Ri}=0$ (g); $\textit{Wi}=5$, $\textit{Ri}=0.143$ (h); $\textit{Wi}=5$, $\textit{Ri}\rightarrow \infty$ (i); $\textit{Wi}=10$, $\textit{Ri}=0$ (j); $\textit{Wi}=10$, $\textit{Ri}=0.143$ (k); and $\textit{Wi}=10$, $\textit{Ri}\rightarrow \infty$ (l).} 
    \label{fig:Isotherms}
\end{figure}

Having explored the flow dynamics, we now delve into the heat transfer characteristics within the current system, focusing on the same combinations of Richardson and Weissenberg numbers as depicted in Fig.~\ref{fig:Streamlines} for the flow dynamics. Figure~\ref{fig:Isotherms} provides a surface plot depicting the non-dimensional temperature distribution and isotherms for various Richardson and Weissenberg number values. In the context of pure forced convection ($Ri = 0$), as presented in the first column of Fig.~\ref{fig:Isotherms}, the isotherms exhibit a pattern akin to concentric circles, resembling the streamlines observed in the flow field. Here, there is a gradual temperature diffusion from the hot inner cylinder towards the outer cylinder, independent of the Weissenberg number's value. It occurs due to the presence of solid body-like rotation within the current flow system, implying that heat transfer primarily will occur through conduction in this case.

On the other hand, in the case of pure free convection (with $Ri \rightarrow \infty$), as depicted in the last column of Fig.~\ref{fig:Isotherms}, the isotherms exhibit more pronounced distortion compared to the case of pure forced convection. This distortion arises due to the fluid's recirculation along the perimeter of the hot inner cylinder towards the outer cold cylinder. The isotherms are notably concentrated in the vicinity of the hot inner cylinder, indicating the presence of a thin thermal boundary layer in this region. At the top gap region, located at $\theta = \pi/2$ between the two cylinders (sub-Fig.~\ref{fig:Isotherms}(c)), the isotherms adopt a mushroom-like shape. At $Wi = 1$, they tilt towards the right-hand side of the origin due to the flow's asymmetry at this Weissenberg number, as evidenced in sub-Fig.~\ref{fig:Isotherms}(f). With an increase in the Weissenberg number, such as the results at $Wi = 10$ illustrated in sub-Fig.~\ref{fig:Isotherms}(l), the thickness of these mushroom-shaped isotherms at the top gap region expands, accompanied by enhanced mixing of hot and cold fluids. Regardless of the Weissenberg number's value, the fluids exhibit reduced mixing at the bottom gap region positioned at $\theta = 3\pi/2$ between the two cylinders.       

\begin{figure}[hbt!]
    \centering
    \includegraphics[trim=0cm 0cm 0cm 0cm,clip,width=13cm]{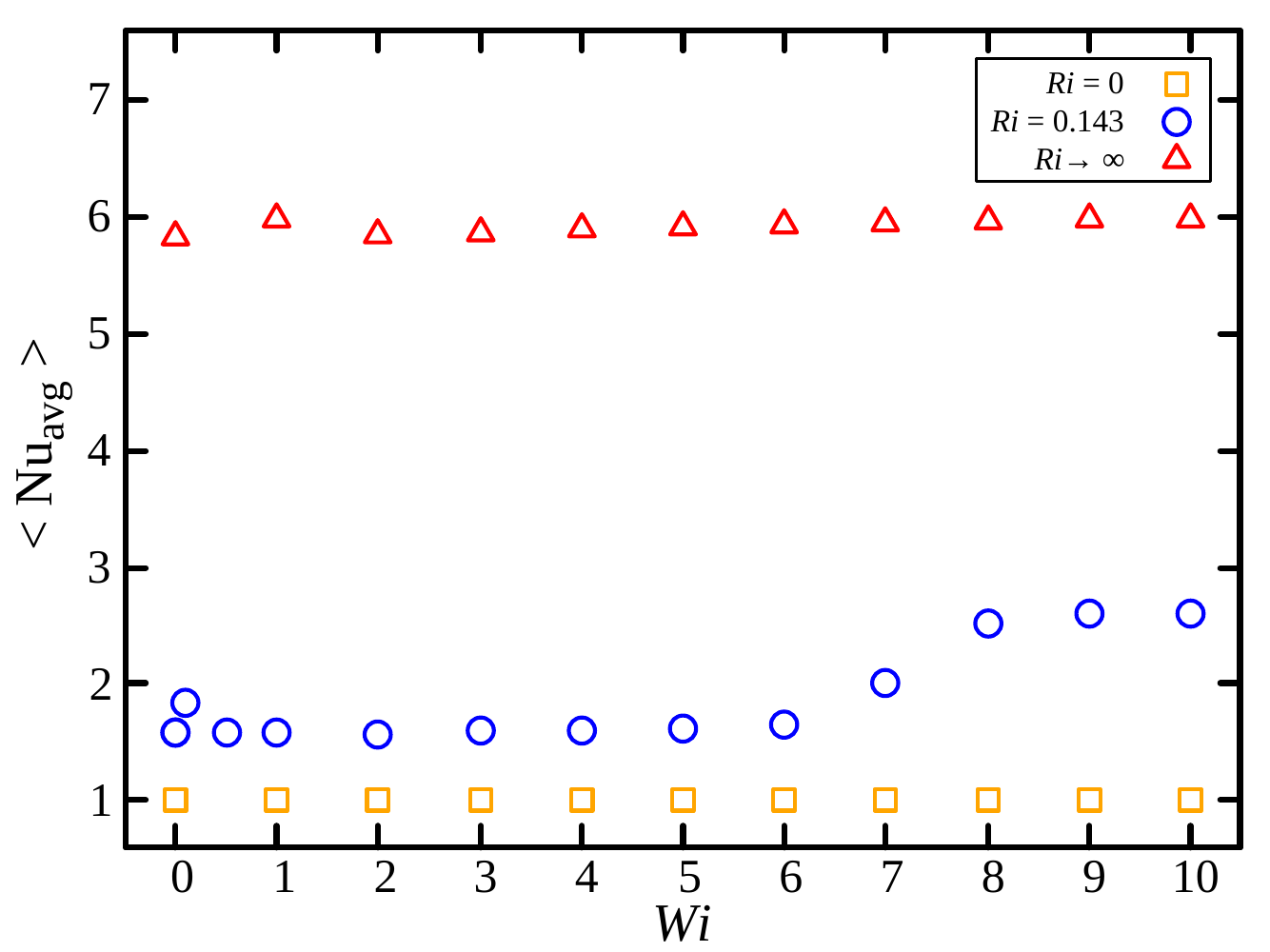}
    \caption{Variation of the surface and time-averaged Nusselt number calculated on the surface of the inner cylinder with Weissenberg number for all three modes of heat transfer, namely, forced ($\textit{Ri}=0$), mixed ($\textit{Ri}=0.143$), and natural ($\textit{Ri} \rightarrow \infty$) convection.} 
    \label{fig:Nu}
\end{figure}

In contrast, during mixed convection heat transfer, the isotherms near the hot cylinder adopt a circular shape akin to the cylinder itself. However, as we move away from the inner cylinder, the isotherms exhibit distortion, which becomes particularly pronounced at $Wi = 10$ (sub-Fig.~\ref{fig:Isotherms}(k)). This distortion is a result of elastic turbulence-induced convection within the system. It is worth noting that the isotherms become increasingly concentrated, and the thermal boundary layer gradually diminishes as the Weissenberg number increases. Consequently, one would anticipate an augmentation in the heat transfer rate with an increase in the Weissenberg number during mixed convection. This effect will now be quantitatively examined through the variation of the Nusselt number, calculated along the surface of the hot inner cylinder. The local value of this Nusselt number, denoted as $(Nu_l)$, at a point on the inner cylinder's surface is estimated as follows:
\begin{equation}
   Nu_l = \frac{h_l \sigma}{k} = - \left (\frac{\partial \phi}{\partial \bm{n}_s} \right)_{surface}
\end{equation}
where $h_l$ is the local heat transfer coefficient and $\bm{n}_s$ is the unit normal vector drawn on the surface of the inner cylinder. Furthermore, the surface average values $(Nu_{avg})$ of this local Nusselt number are obtained by integrating the local values over the whole surface of the inner cylinder as follows:
\begin{equation}
   Nu_{avg} = \frac{h \sigma}{k} = \frac{1}{S} \:\int_{S}Nu_l\:dS
\end{equation}
where $h$ is the average heat transfer coefficient. Figure~\ref{fig:Nu} illustrates how the time-averaged values of the surface-averaged Nusselt number ($< Nu_{avg} >$) vary with the Weissenberg number at different Richardson number values. Notably, during forced convective heat transfer, $< Nu_{avg} >$ remains nearly constant at around one across the entire range of Weissenberg numbers investigated in this study. This value of the average Nusselt number suggests that heat transfer predominantly occurs via conduction in this case, with limited dependence on the fluid's kinematic properties, such as viscosity and relaxation time. Consequently, the heat transfer rate remains consistent and is not dependent upon the Weissenberg number.

In contrast, in the case of pure free convection ($Ri \rightarrow \infty$), the heat transfer rate increases due to the presence of buoyancy-induced convection within the system. This convection facilitates mixing hot fluid near the inner cylinder with cold fluid near the outer cylinder, enhancing heat transfer. As a result, the average Nusselt number increases by approximately sixfold compared to forced convection. However, once again, the Nusselt number values show minimal dependence on the Weissenberg number in this mode of heat transfer. This is likely because, while buoyancy-induced convection currents are generated in this mode, their strength is comparatively lower than that observed in forced or mixed convection. This is evident in Fig.~\ref{fig:Streamlines}, which presents the velocity magnitude for different heat transfer modes. This lower strength of buoyancy-induced convection currents results in less stretching of polymer molecules and generates lesser elastic stresses, which are insufficient to transit the flow field into the regime of severe elastic turbulence, where a significant change in heat transfer rate with the Weissenberg number would be expected. 

On the other hand, in the mixed convection case, a different trend is observed in the variation of the average Nusselt number. In this mode of heat transfer, the average Nusselt number value is much lower than the value seen for free convection but higher than forced convection. It remains constant to a value of around 1.6 up to $Wi = 6$, and then it starts to increase with the Weissenberg number, and ultimately, it again becomes constant at higher values of the Weissenberg number. For instance, the Nusselt number value is increased by almost 64\%  in viscoelastic fluids with $Wi = 10$, compared to Newtonian fluids under identical conditions. This enhancement in the heat transfer rate at high values of the Weissenberg number for viscoelastic fluids is due to the elastic instability-induced increased chaotic convection within the present system, as seen from Fig.~\ref{fig:Streamlines}. This results in the decrease of the thermal boundary layer and an increase in the mixing of hot and cold fluids as observed from isotherm plot presented in Fig.~\ref{fig:Isotherms}. The flow dynamics and heat transfer aspects, particularly the enhancement in the heat transfer rate at high values of the Weissenberg number in the case of mixed convection, are further illustrated with the help of the viscoelastic kinetic energy budget analysis presented in the subsequent section.               

\subsection{Viscoelastic kinetic energy budget analysis}

Based on the analysis of Zheng et al.~\cite{zheng2023time} and Cheng et al.~\cite{cheng2017effect}, we have performed the viscoelastic kinetic energy budget analysis to gain more insights into the fluctuating and time-dependent flow and heat transfer phenomena seen in the mixed convection heat transfer. This analysis will facilitate a better understanding of the underlying physics considering the different components of fluid field energy, which denote the global and local kinetic energy exchanges, as shown in the following equation. 

\begin{equation} \label{eq:VKEBA}
     \frac{dE_{ij}}{dt} = {\chi_{P,ij}} + {\chi_{D,ij}} + {\chi_{V,ij}} + {\chi_{G,ij}} + {\chi_{F,ij}}     
\end{equation}

In the above equation, $E_{ij} = \frac{1}{2} <u_i> <u_j> \delta_{ij}$ is the kinetic energy, $\chi_{P,ij} = - <u_i> <u_j> \frac{\partial <u_i>}{\partial x_j} $ accounts for the inertial energy production, $\chi_{D,ij} = - \frac{\partial (<p><u_i>)}{\partial x_j} \delta_{ij} + \frac{\beta}{Re} \frac{\partial^2 E_{ij}}{\partial x^2_j}$ signifies the energy transport due to pressure diffusion and molecular viscous transport, $\chi_{V,ij} = - \frac{\beta}{Re} \frac{\partial <u_i>}{\partial x_j} \frac{\partial <u_i>}{\partial x_j}$ represents the viscous dissipation term, $\chi_{G,ij} = (\frac{1-\beta}{Re}) \left(\frac{\partial (<u_i><\tau_{ij}>)}{\partial x_j} - <\tau_{ij}> \frac{\partial <u_i>}{\partial x_j}\right)$ denotes the energy transition between the flow structure and polymer molecules due to their stretching and relaxation mechanisms, and $\chi_{F,ij} = \frac{Gr}{Re^2} <\phi> <u_i> \delta_{i2}$ represents the buoyancy flux input. Here, $\delta_{ij}$ is the Kronecker delta function, and $<.>$ operator denotes the time-averaged value of a variable. Furthermore, a spatial average over the whole computational domain was also performed for each component as follows: $E = \frac{1}{S} \:\int \int E_{ij}\:dx_1 dx_2$, $\chi_{P} = \frac{1}{S} \:\int \int \chi_{P,ij}\:dx_1 dx_2$, $\chi_{D} = \frac{1}{S} \:\int \int \chi_{D,ij}\:dx_1 dx_2$, $\chi_{V} = \frac{1}{S} \:\int \int \chi_{V,ij}\:dx_1 dx_2$, $\chi_{G} = \frac{1}{S} \:\int \int \chi_{G,ij}\:dx_1 dx_2$, and $\chi_{F} = \frac{1}{S} \:\int \int \chi_{F,ij}\:dx_1 dx_2$, where $S$ is the surface area of the whole computational domain.

\begin{figure}
    \centering
    \includegraphics[trim=0cm 0cm 0cm 0cm,clip,width=11cm]{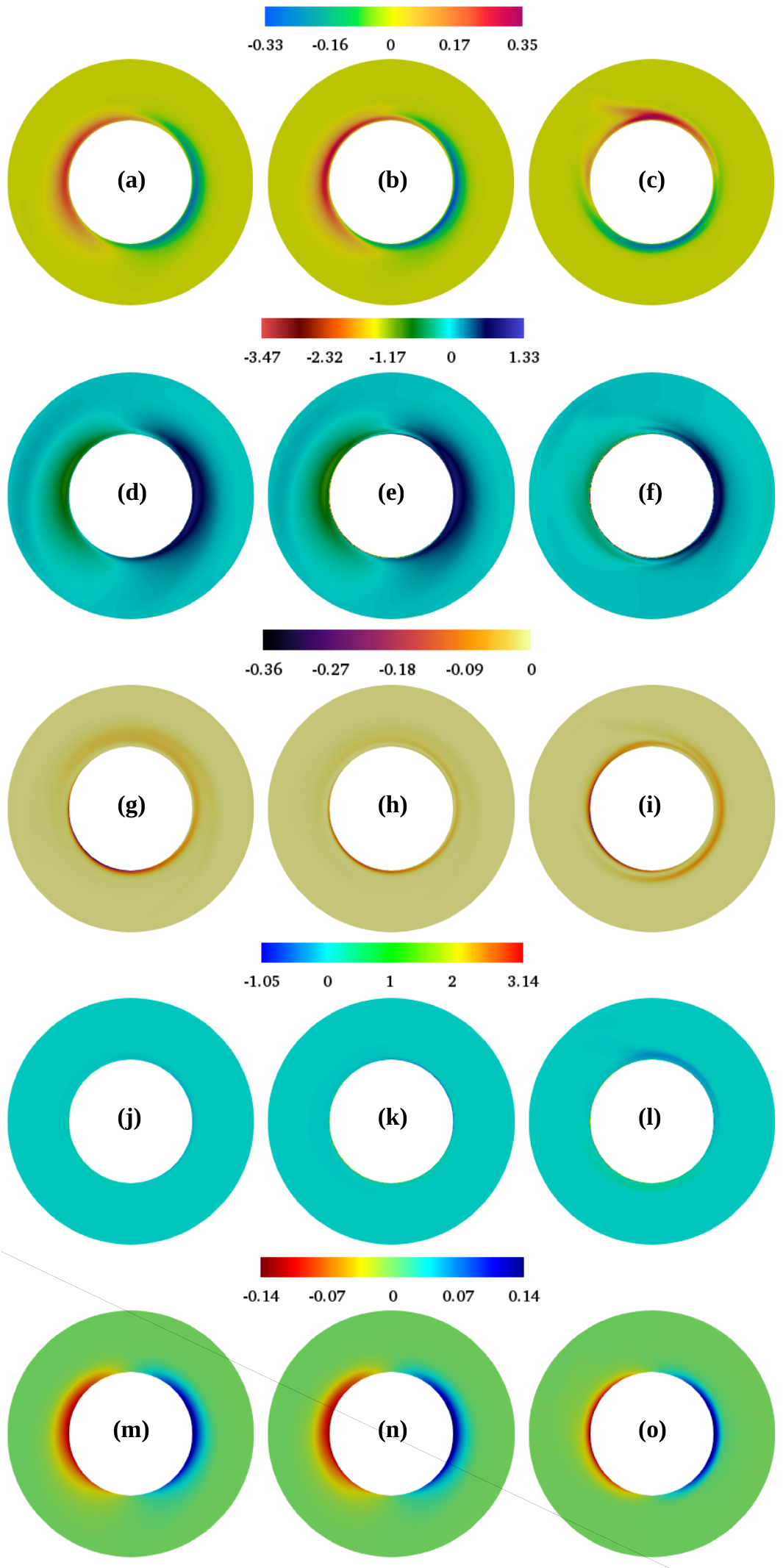}
    \caption{Spatial variation of time-averaged terms appearing in the viscoelastic kinetic budget analysis (Eq.~\ref{eq:VKEBA}) with Weissenberg number at $Ra=10^6$ and $Ri=0.143$. The combinations shown here are for: $\chi_{P,ij}$, $\textit{Wi}=0$ (a); $\chi_{P,ij}$, $\textit{Wi}=5$ (b); $\chi_{P,ij}$, $\textit{Wi}=10$ (c); $\chi_{D,ij}$, $\textit{Wi}=0$ (d); $\chi_{D,ij}$, $\textit{Wi}=5$ (e); $\chi_{D,ij}$, $\textit{Wi}=10$ (f); $\chi_{V,ij}$, $\textit{Wi}=0$ (g); $\chi_{V,ij}$, $\textit{Wi}=5$ (h); $\chi_{V,ij}$, $\textit{Wi}=10$ (i); $\chi_{G,ij}$, $\textit{Wi}=0.01$ (j); $\chi_{G,ij}$, $\textit{Wi}=5$ (k); $\chi_{G,ij}$, $\textit{Wi}=10$ (l); $\chi_{F,ij}$, $\textit{Wi}=0$ (m); $\chi_{F,ij}$, $\textit{Wi}=5$ (n); and $\chi_{F,ij}$, $\textit{Wi}=10$ (o).} 
    \label{fig:All_chi}
\end{figure}

In sub-Fig.~\ref{fig:All_chi}((a)-(c)), we illustrate the inertial energy production in viscoelastic fluids with varying Weissenberg numbers, including the case of a Newtonian fluid for which the Weissenberg number is zero. Irrespective of the fluid type, whether Newtonian or viscoelastic, we observe consistently high inertial energy production near the inner cylinder. This is attributed to the substantial impact of inner cylinder rotation in this region. As the Weissenberg number increases, the width of the high $\chi_{P}$ region decreases. Notably, at higher Weissenberg numbers, such as at $Wi = 10$ (see sub-Figs.~\ref{fig:All_chi}(c)), there is a significant alteration in the positions of the highest (positive) and lowest (negative) regions of inertial energy production. In this case, these regions become oriented horizontally to each other, whereas, for Newtonian and viscoelastic fluids with lower Weissenberg numbers (e.g., sub-Fig.~\ref{fig:All_chi}(a)), they are oriented vertically to each other. This observation suggests that the flow field undergoes substantial changes at high Weissenberg numbers, subsequently exerting a significant influence on the heat transfer rate.

Sub-Figs.~\ref{fig:All_chi}((d)-(f)) present the spatial distribution of energy transport arising from pressure diffusion and molecular viscous transport. In the case of Newtonian fluids (sub-Fig.~\ref{fig:All_chi}(d)) and viscoelastic fluids with low Weissenberg numbers, the dominant mechanism for energy transport is molecular viscous transport, particularly near the surface of the inner rotating cylinder. However, as the Weissenberg number gradually increases to higher values, the energy transport due to pressure diffusion surpasses that of molecular viscous transport. This shift is anticipated, given that fluid viscoelasticity leads to an increased pressure gradient~\cite{james2023pressure, bendova2009pressure}, consequently enhancing the contribution of pressure diffusion to energy transport. Similar to the trend observed in inertial energy production, the region characterized by high $\chi_{D}$ also contracts as the Weissenberg number increases. Sub-Figs.~\ref{fig:All_chi}((g)-(i)) illustrate the spatial distribution of viscoelastic kinetic energy transfer attributed to viscous dissipation for different Weissenberg number values. Irrespective of the fluid type, this energy transfer remains close to zero except in the vicinity of the inner rotating cylinder, where intense shearing between fluid layers occurs. Notably, this transfer is more pronounced in the case of Newtonian fluids (sub-Fig.\ref{fig:All_chi}(g)) and high Weissenberg number viscoelastic fluids (sub-Fig.\ref{fig:All_chi}(i)). Sub-Figs.~\ref{fig:All_chi}((j)-(l)) presents the spatial variation of energy exchange between flow structures and polymer molecules. When this term is positive, it acts as an energy source in the system, while it functions as an energy sink when negative~\cite{cheng2017effect}. As the Weissenberg number increases gradually, both contributions intensify. Specifically, at the top of the inner cylinder, energy is transferred from the flow structure to the polymer molecules, while at the bottom side of the inner cylinder, the energy flows from the polymers to the flow field. On the other hand, sub-Fig.~\ref{fig:All_chi}((m)-(o)) demonstrates the spatial variation of kinetic energy transfer due to buoyant flux. This transfer is prominent along the two vertical sides of the inner cylinder, and its width diminishes as the Weissenberg number increases.

\begin{figure}
    \centering
    \includegraphics[trim=0cm 0cm 0cm 0cm,clip,width=13cm]{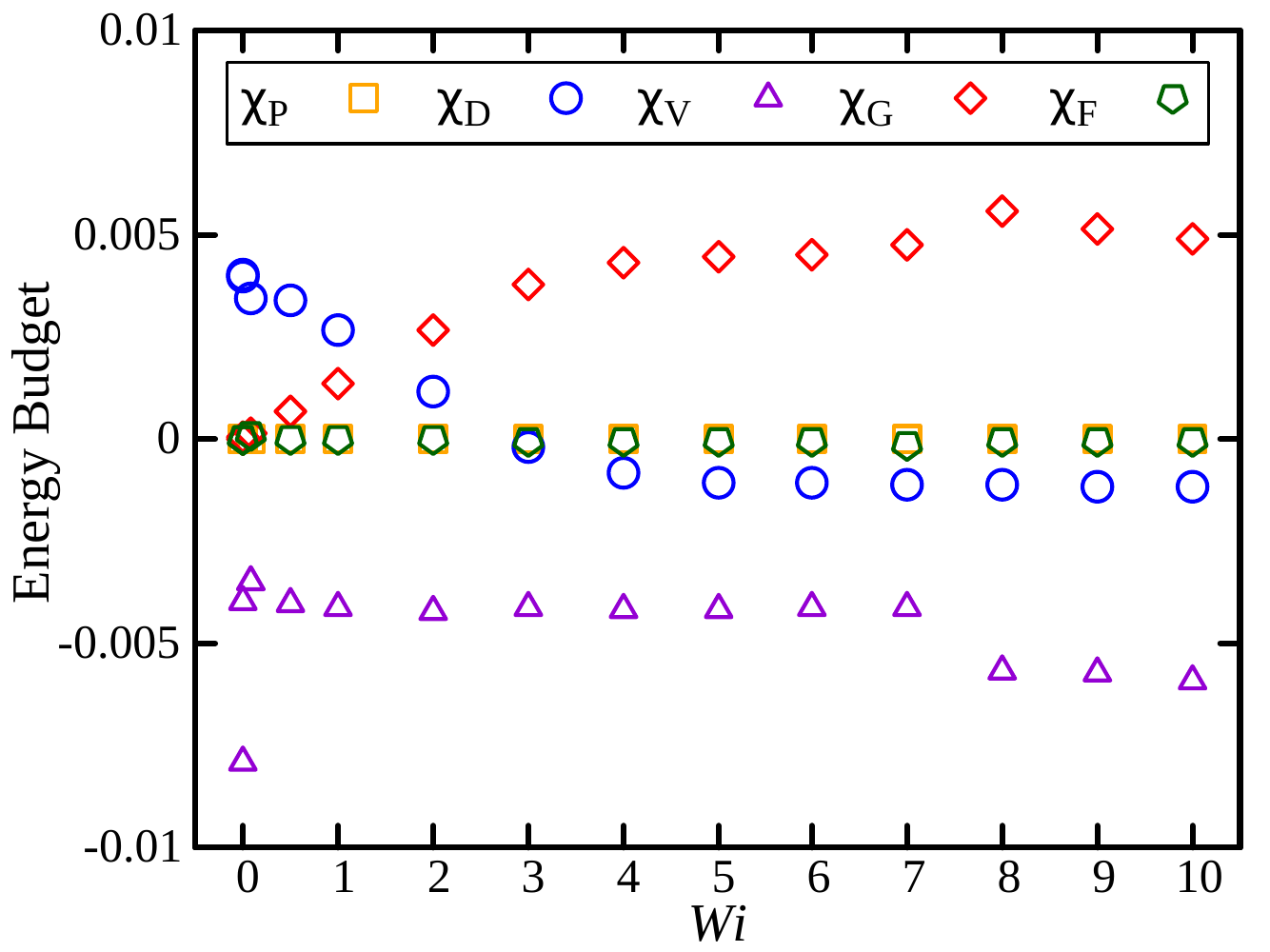}
    \caption{Variation of different terms (tempo-spatial averaged) appearing in the viscoelastic kinetic energy budget analysis with Weissenberg number for $Ra=10^6$ and $Ri=0.143$.} 
    \label{fig:EnergyBudget}
\end{figure}

The spatial variations of different components of kinetic energy transfer are depicted in time-averaged surface plots, providing insights into their distribution. However, for a comprehensive analysis of heat transfer aspects, it is crucial to calculate and understand the surface-averaged values. Thus, in Fig.~\ref{fig:EnergyBudget}, we present the variations of various components of the kinetic energy budget with respect to the Weissenberg number. The surface-averaged value of inertial energy production $(\chi_{P})$ remains consistently low, near zero, and largely independent of the Weissenberg number. In contrast, energy transfer attributed to pressure diffusion and molecular viscous transport $(\chi_{D})$ exhibits variations. It is highest for Newtonian fluids and gradually decreases for viscoelastic fluids as the Weissenberg number increases up to around 5. Beyond this point, it remains relatively constant. As mentioned earlier, for Newtonian and low Weissenberg number viscoelastic fluids, energy transfer due to viscous molecular transport (which contributes positively) dominates. However, as the Weissenberg number increases, energy transport due to pressure diffusion (with a negative contribution) becomes dominant, reducing the overall energy contribution. The energy transfer $(\chi_{V})$ attributed to viscous dissipation (which results in energy loss) is highest in magnitude for Newtonian fluids due to the sole presence of solvent molecules, promoting shearing among them. In viscoelastic fluids, the introduction of polymer molecules reduces the solvent's concentration, leading to a decrease in the viscous dissipation associated with shearing among solvent fluid layers. This magnitude remains relatively constant up to a Weissenberg number of approximately 7. At higher Weissenberg numbers, the magnitude of viscous dissipation increases, likely due to the flow field becoming chaotic and fluctuating due to elastic instability. This chaotic motion enhances shearing among solvent fluid layers, subsequently increasing viscous dissipation. The buoyancy flux input $(\chi_{F})$ remains consistently small, close to zero, and independent of the Weissenberg number. Conversely, the surface-averaged value of energy exchange between fluid structure and polymer molecules $(\chi_{G})$ remains positive and rises with increasing Weissenberg number. This observation suggests that energy transfers from polymer molecules to the fluid structure, leading to increased chaotic motion within the system. As discussed in the preceding subsection, it eventually contributes to an increase in the heat transfer rate. Our observation and explanation are in line with that seen and provided by Cheng et al.~\cite{cheng2017effect} for the enhancement of heat transfer rate with the Weissenberg number during the Rayleigh-Bennard convection of viscoelastic fluids in a square cavity.

\section{\label{sec:5}Conclusions}
This study conducted an extensive numerical investigation to explore forced, free, and mixed convection heat transfer phenomena in viscoelastic fluids confined between two concentric cylinders, with the inner cylinder undergoing rotation. The rheological behavior of the viscoelastic fluid was represented using the Oldroyd-B constitutive equation, and the numerical simulations were executed using the open-source code OpenFOAM to solve the governing equations. Forced convection conditions maintained a steady state within the range of parameters examined in this study. However, as the viscoelasticity of the fluid, quantified by the Weissenberg number, increased, a transition in the flow behavior was observed in the cases of free and mixed convection conditions. In particular, the flow field shifted from a steady state to unsteady periodic, quasi-periodic, and ultimately aperiodic or chaotic states. This transition was attributed to the emergence of elastic instability and ensuing elastic turbulence within the flow field as fluid viscoelasticity increased. Regarding heat transfer, the Weissenberg number exhibited minimal influence on heat transfer rates in the case of forced and free convection. However, in the case of mixed convection, heat transfer rates demonstrated an increase with the Weissenberg number. A comprehensive analysis of the viscoelastic kinetic energy budget was conducted to elucidate this heat transfer enhancement. The investigation revealed a notable phenomenon wherein energy transition occurred from the polymers to the flow field within viscoelastic fluids. This energy transfer intensified with the Weissenberg number, consequently augmenting the system's kinetic energy. This increased kinetic energy generated chaotic motion within the system and correspondingly increased the heat transfer rate.

\begin{acknowledgments}
We would like to thank Mr. M. Kumar for making the geometry and the National Supercomputing Mission (NSM) for providing computing resources in `PARAM Himalaya' at IIT Mandi, which is implemented by C-DAC and supported by the Ministry of Electronics and Information Technology (MeitY) and Department of Science and Technology (DST), Government of India.
\end{acknowledgments}

\appendix

\section{\label{App}Governing equations for purely free convection}
We non-dimensionalize all the governing equations (Eqs.~\ref{eq:continuity}--~\ref{eq:Conformation_transport}) using the following scaling variables in the case of purely free convection:

$x_i = \frac{x^{*}_i}{\sigma}$, $u_i = \frac{u^{*}_i}{\sqrt{\sigma g \beta_T \Delta T}}$, $t = t^{*} \sqrt{\frac{g \beta_T \Delta T}{\sigma}}$, $p = \frac{p^{*}}{\rho_{ref} \sigma g \beta_T \Delta T}$, $\tau^{p}_{ij}=\frac{\tau^{p*}_{ij}}{\eta_p}\sqrt{\frac{\sigma}{g \beta_T \Delta T}}$ and $\phi = \frac{T-T_C}{T_H-T_C}$. 
\\ 
The non-dimensional forms of the governing equations are obtained as follows:

Continuity equation:
\begin{equation}
   \frac{\partial {u_i}}{\partial x_i} = 0
\end{equation}

Momentum equation: 
\begin{equation}
     \frac{\partial {u_i}}{\partial t} + {u_j} \frac{\partial {u_i}}{\partial x_j} = - \frac{\partial p}{\partial x_i} + \frac{\beta}{\sqrt{Ra/Pr}} \frac{\partial}{\partial x_j}\left (\frac{\partial {u_i}}{\partial x_j} \right)+\frac{1-{\beta}}{\sqrt{Ra/Pr}}\frac{\partial {\tau^p_{ij}}}{\partial x_j} + \phi \delta_{i2}
\end{equation}

Energy equation: 
\begin{equation}
     \frac{\partial \phi}{\partial t} + {u_j} \frac{\partial \phi}{\partial x_j} = \frac{1}{\sqrt{Ra Pr}}\frac{\partial}{\partial x_j}\left (\frac{\partial \phi}{\partial x_j} \right)
\end{equation}

Polymeric conformation tensor transport equation:
\begin{equation}
       \dfrac{\partial{{C}_{ij}}}{\partial{t}}+{u}_{k}\frac{\partial {C}_{ij}}{\partial x_k}-\frac{\partial {u}_{i}}{\partial x_k}{C}_{kj}-\frac{\partial {u}_{j}}{\partial x_k}{C}_{ik}=\frac{\delta_{ij}-{C}_{ij}}{Wi}
\end{equation}
In the above equations, $Ra = \frac{{C_p \rho}^2 g \beta_T (T_H-T_C) {\sigma}^3}{{\eta_0} {k}}$ is the Rayleigh number, $Pr = \frac{C_p \eta_0}{k}$ is the Prandtl number, $Wi = \lambda \sqrt{\frac{g \beta_T \Delta T}{\sigma}}$ is the Weissenberg number, and $\beta = \frac{\eta_s}{\eta_0} = \frac{\eta_s}{\eta_s + \eta_p}$ is the polymer viscosity ratio, which is defined as the ratio of the solvent to that of the zero-shear rate viscosity of the polymer solution.


\bibliography{References}

\begin{thebibliography}{61}%
\makeatletter
\providecommand \@ifxundefined [1]{%
 \@ifx{#1\undefined}
}%
\providecommand \@ifnum [1]{%
 \ifnum #1\expandafter \@firstoftwo
 \else \expandafter \@secondoftwo
 \fi
}%
\providecommand \@ifx [1]{%
 \ifx #1\expandafter \@firstoftwo
 \else \expandafter \@secondoftwo
 \fi
}%
\providecommand \natexlab [1]{#1}%
\providecommand \enquote  [1]{``#1''}%
\providecommand \bibnamefont  [1]{#1}%
\providecommand \bibfnamefont [1]{#1}%
\providecommand \citenamefont [1]{#1}%
\providecommand \href@noop [0]{\@secondoftwo}%
\providecommand \href [0]{\begingroup \@sanitize@url \@href}%
\providecommand \@href[1]{\@@startlink{#1}\@@href}%
\providecommand \@@href[1]{\endgroup#1\@@endlink}%
\providecommand \@sanitize@url [0]{\catcode `\\12\catcode `\$12\catcode
  `\&12\catcode `\#12\catcode `\^12\catcode `\_12\catcode `\%12\relax}%
\providecommand \@@startlink[1]{}%
\providecommand \@@endlink[0]{}%
\providecommand \url  [0]{\begingroup\@sanitize@url \@url }%
\providecommand \@url [1]{\endgroup\@href {#1}{\urlprefix }}%
\providecommand \urlprefix  [0]{URL }%
\providecommand \Eprint [0]{\href }%
\providecommand \doibase [0]{http://dx.doi.org/}%
\providecommand \selectlanguage [0]{\@gobble}%
\providecommand \bibinfo  [0]{\@secondoftwo}%
\providecommand \bibfield  [0]{\@secondoftwo}%
\providecommand \translation [1]{[#1]}%
\providecommand \BibitemOpen [0]{}%
\providecommand \bibitemStop [0]{}%
\providecommand \bibitemNoStop [0]{.\EOS\space}%
\providecommand \EOS [0]{\spacefactor3000\relax}%
\providecommand \BibitemShut  [1]{\csname bibitem#1\endcsname}%
\let\auto@bib@innerbib\@empty
\bibitem [{\citenamefont {Gazley~Jr}(1958)}]{gazley1958heat}%
  \BibitemOpen
  \bibfield  {author} {\bibinfo {author} {\bibfnamefont {Carl}\ \bibnamefont
  {Gazley~Jr}},\ }\bibfield  {title} {\enquote {\bibinfo {title} {Heat-transfer
  characteristics of the rotational and axial flow between concentric
  cylinders},}\ }\href@noop {} {\bibfield  {journal} {\bibinfo  {journal}
  {Trans. American Soc. Mech. Eng.}\ }\textbf {\bibinfo {volume} {80}},\
  \bibinfo {pages} {79--89} (\bibinfo {year} {1958})}\BibitemShut {NoStop}%
\bibitem [{\citenamefont {Bjorklund}\ and\ \citenamefont
  {Kays}(1959)}]{bjorklund1959heat}%
  \BibitemOpen
  \bibfield  {author} {\bibinfo {author} {\bibfnamefont {IS}~\bibnamefont
  {Bjorklund}}\ and\ \bibinfo {author} {\bibfnamefont {WM}~\bibnamefont
  {Kays}},\ }\bibfield  {title} {\enquote {\bibinfo {title} {Heat transfer
  between concentric rotating cylinders},}\ }\href@noop {} {\bibfield
  {journal} {\bibinfo  {journal} {J. Heat Transfer}\ }\textbf {\bibinfo
  {volume} {81}},\ \bibinfo {pages} {175--183} (\bibinfo {year}
  {1959})}\BibitemShut {NoStop}%
\bibitem [{\citenamefont {Taylor}(1923)}]{taylor}%
  \BibitemOpen
  \bibfield  {author} {\bibinfo {author} {\bibfnamefont {G.~I.}\ \bibnamefont
  {Taylor}},\ }\bibfield  {title} {\enquote {\bibinfo {title} {Stability of a
  viscous liquid contained between two rotating cylinders},}\ }\href@noop {}
  {\bibfield  {journal} {\bibinfo  {journal} {Philos. Trans. R . Soc. London
  Ser. A}\ }\textbf {\bibinfo {volume} {223}},\ \bibinfo {pages} {289--343}
  (\bibinfo {year} {1923})}\BibitemShut {NoStop}%
\bibitem [{\citenamefont {Di~Prima}\ and\ \citenamefont
  {Swinney}(2005)}]{di2005instabilities}%
  \BibitemOpen
  \bibfield  {author} {\bibinfo {author} {\bibfnamefont {RC}~\bibnamefont
  {Di~Prima}}\ and\ \bibinfo {author} {\bibfnamefont {Harry~L}\ \bibnamefont
  {Swinney}},\ }\bibfield  {title} {\enquote {\bibinfo {title} {Instabilities
  and transition in flow between concentric rotating cylinders},}\ }\href@noop
  {} {\bibfield  {journal} {\bibinfo  {journal} {Hydrodynamic instabilities and
  the transition to turbulence}\ ,\ \bibinfo {pages} {139--180}} (\bibinfo
  {year} {2005})}\BibitemShut {NoStop}%
\bibitem [{\citenamefont {Coles}(1965)}]{coles}%
  \BibitemOpen
  \bibfield  {author} {\bibinfo {author} {\bibfnamefont {D.}~\bibnamefont
  {Coles}},\ }\bibfield  {title} {\enquote {\bibinfo {title} {Transition in
  circular couette flow},}\ }\href@noop {} {\bibfield  {journal} {\bibinfo
  {journal} {J. Fluid Mech.}\ }\textbf {\bibinfo {volume} {21 (3)}},\ \bibinfo
  {pages} {385--425} (\bibinfo {year} {1965})}\BibitemShut {NoStop}%
\bibitem [{\citenamefont {F{\'e}not}\ \emph {et~al.}(2011)\citenamefont
  {F{\'e}not}, \citenamefont {Bertin}, \citenamefont {Dorignac},\ and\
  \citenamefont {Lalizel}}]{fenot2011review}%
  \BibitemOpen
  \bibfield  {author} {\bibinfo {author} {\bibfnamefont {M}~\bibnamefont
  {F{\'e}not}}, \bibinfo {author} {\bibfnamefont {Y}~\bibnamefont {Bertin}},
  \bibinfo {author} {\bibfnamefont {E}~\bibnamefont {Dorignac}}, \ and\
  \bibinfo {author} {\bibfnamefont {G}~\bibnamefont {Lalizel}},\ }\bibfield
  {title} {\enquote {\bibinfo {title} {A review of heat transfer between
  concentric rotating cylinders with or without axial flow},}\ }\href@noop {}
  {\bibfield  {journal} {\bibinfo  {journal} {Int. J. Therm. Sci.}\ }\textbf
  {\bibinfo {volume} {50}},\ \bibinfo {pages} {1138--1155} (\bibinfo {year}
  {2011})}\BibitemShut {NoStop}%
\bibitem [{\citenamefont {Kuehn}\ and\ \citenamefont
  {Goldstein}(1976)}]{kuehn}%
  \BibitemOpen
  \bibfield  {author} {\bibinfo {author} {\bibfnamefont {T.~H.}\ \bibnamefont
  {Kuehn}}\ and\ \bibinfo {author} {\bibfnamefont {R.~J.}\ \bibnamefont
  {Goldstein}},\ }\bibfield  {title} {\enquote {\bibinfo {title} {An
  experimental and theoretical study of natural convection in the annulus
  between horizontal concentric cylinders},}\ }\href@noop {} {\bibfield
  {journal} {\bibinfo  {journal} {J. Fluid Mech.}\ }\textbf {\bibinfo {volume}
  {74}},\ \bibinfo {pages} {695--719} (\bibinfo {year} {1976})}\BibitemShut
  {NoStop}%
\bibitem [{\citenamefont {Kumar}(1988)}]{kumar1988study}%
  \BibitemOpen
  \bibfield  {author} {\bibinfo {author} {\bibfnamefont {Ranganathan}\
  \bibnamefont {Kumar}},\ }\bibfield  {title} {\enquote {\bibinfo {title}
  {Study of natural convection in horizontal annuli},}\ }\href@noop {}
  {\bibfield  {journal} {\bibinfo  {journal} {Int. J. Heat Mass Transfer}\
  }\textbf {\bibinfo {volume} {31}},\ \bibinfo {pages} {1137--1148} (\bibinfo
  {year} {1988})}\BibitemShut {NoStop}%
\bibitem [{\citenamefont {Yoo}(1999)}]{yoo1999prandtl}%
  \BibitemOpen
  \bibfield  {author} {\bibinfo {author} {\bibfnamefont {Joo-Sik}\ \bibnamefont
  {Yoo}},\ }\bibfield  {title} {\enquote {\bibinfo {title} {Prandtl number
  effect on bifurcation and dual solutions in natural convection in a
  horizontal annulus},}\ }\href@noop {} {\bibfield  {journal} {\bibinfo
  {journal} {Int. J. Heat Mass Transfer}\ }\textbf {\bibinfo {volume} {42}},\
  \bibinfo {pages} {3279--3290} (\bibinfo {year} {1999})}\BibitemShut {NoStop}%
\bibitem [{\citenamefont {Desai}\ and\ \citenamefont
  {Vafai}(1994)}]{desai1994investigation}%
  \BibitemOpen
  \bibfield  {author} {\bibinfo {author} {\bibfnamefont {CP}~\bibnamefont
  {Desai}}\ and\ \bibinfo {author} {\bibfnamefont {K}~\bibnamefont {Vafai}},\
  }\bibfield  {title} {\enquote {\bibinfo {title} {An investigation and
  comparative analysis of two-and three-dimensional turbulent natural
  convection in a horizontal annulus},}\ }\href@noop {} {\bibfield  {journal}
  {\bibinfo  {journal} {Int. J. Heat Mass Transfer}\ }\textbf {\bibinfo
  {volume} {37}},\ \bibinfo {pages} {2475--2504} (\bibinfo {year}
  {1994})}\BibitemShut {NoStop}%
\bibitem [{\citenamefont {Khanafer}\ and\ \citenamefont
  {Chamkha}(2003)}]{khanafer}%
  \BibitemOpen
  \bibfield  {author} {\bibinfo {author} {\bibfnamefont {K.}~\bibnamefont
  {Khanafer}}\ and\ \bibinfo {author} {\bibfnamefont {A.~J.}\ \bibnamefont
  {Chamkha}},\ }\bibfield  {title} {\enquote {\bibinfo {title} {Mixed
  convection within a porous heat generating horizontal annulus},}\ }\href@noop
  {} {\bibfield  {journal} {\bibinfo  {journal} {Int. J. Heat Mass Transfer}\
  }\textbf {\bibinfo {volume} {46}},\ \bibinfo {pages} {1725--1735} (\bibinfo
  {year} {2003})}\BibitemShut {NoStop}%
\bibitem [{\citenamefont {Nada}\ and\ \citenamefont
  {Said}(2019)}]{nada2019effects}%
  \BibitemOpen
  \bibfield  {author} {\bibinfo {author} {\bibfnamefont {SA}~\bibnamefont
  {Nada}}\ and\ \bibinfo {author} {\bibfnamefont {MA}~\bibnamefont {Said}},\
  }\bibfield  {title} {\enquote {\bibinfo {title} {Effects of fins geometries,
  arrangements, dimensions and numbers on natural convection heat transfer
  characteristics in finned-horizontal annulus},}\ }\href@noop {} {\bibfield
  {journal} {\bibinfo  {journal} {Int. J. Thermal Sci.}\ }\textbf {\bibinfo
  {volume} {137}},\ \bibinfo {pages} {121--137} (\bibinfo {year}
  {2019})}\BibitemShut {NoStop}%
\bibitem [{\citenamefont {Powe}\ \emph {et~al.}(1969)\citenamefont {Powe},
  \citenamefont {Carley},\ and\ \citenamefont {Bishop}}]{powe}%
  \BibitemOpen
  \bibfield  {author} {\bibinfo {author} {\bibfnamefont {R.~E.}\ \bibnamefont
  {Powe}}, \bibinfo {author} {\bibfnamefont {C.~T.}\ \bibnamefont {Carley}}, \
  and\ \bibinfo {author} {\bibfnamefont {E.~H.}\ \bibnamefont {Bishop}},\
  }\bibfield  {title} {\enquote {\bibinfo {title} {Free convective flow
  patterns in cylindrical annuli},}\ }\href@noop {} {\bibfield  {journal}
  {\bibinfo  {journal} {J. Heat Transfer}\ }\textbf {\bibinfo {volume} {91}},\
  \bibinfo {pages} {310--314} (\bibinfo {year} {1969})}\BibitemShut {NoStop}%
\bibitem [{\citenamefont {Dawood}\ \emph {et~al.}(2015)\citenamefont {Dawood},
  \citenamefont {Mohammed}, \citenamefont {Sidik}, \citenamefont {Munisamy},\
  and\ \citenamefont {Wahid}}]{dawood2015forced}%
  \BibitemOpen
  \bibfield  {author} {\bibinfo {author} {\bibfnamefont {HK}~\bibnamefont
  {Dawood}}, \bibinfo {author} {\bibfnamefont {HA}~\bibnamefont {Mohammed}},
  \bibinfo {author} {\bibfnamefont {Nor Azwadi~Che}\ \bibnamefont {Sidik}},
  \bibinfo {author} {\bibfnamefont {KM}~\bibnamefont {Munisamy}}, \ and\
  \bibinfo {author} {\bibfnamefont {MA}~\bibnamefont {Wahid}},\ }\bibfield
  {title} {\enquote {\bibinfo {title} {Forced, natural and mixed-convection
  heat transfer and fluid flow in annulus: A review},}\ }\href@noop {}
  {\bibfield  {journal} {\bibinfo  {journal} {Int. Comm. Heat Mass Transfer}\
  }\textbf {\bibinfo {volume} {62}},\ \bibinfo {pages} {45--57} (\bibinfo
  {year} {2015})}\BibitemShut {NoStop}%
\bibitem [{\citenamefont {Aoki}\ \emph {et~al.}(1967)\citenamefont {Aoki},
  \citenamefont {Nohira},\ and\ \citenamefont {Arai}}]{aoki}%
  \BibitemOpen
  \bibfield  {author} {\bibinfo {author} {\bibfnamefont {H.}~\bibnamefont
  {Aoki}}, \bibinfo {author} {\bibfnamefont {H.}~\bibnamefont {Nohira}}, \ and\
  \bibinfo {author} {\bibfnamefont {H.}~\bibnamefont {Arai}},\ }\bibfield
  {title} {\enquote {\bibinfo {title} {Convective heat transfer in an annulus
  with an inner rotating cylinder},}\ }\href@noop {} {\bibfield  {journal}
  {\bibinfo  {journal} {Bull. J. S. M. E}\ }\textbf {\bibinfo {volume}
  {10(39)}},\ \bibinfo {pages} {523--532} (\bibinfo {year} {1967})}\BibitemShut
  {NoStop}%
\bibitem [{\citenamefont {Lee}(1984)}]{lee1}%
  \BibitemOpen
  \bibfield  {author} {\bibinfo {author} {\bibfnamefont {T.~S.}\ \bibnamefont
  {Lee}},\ }\bibfield  {title} {\enquote {\bibinfo {title} {Numerical
  experiments with laminar fluid convection between concentric and eccentric
  heated rotating cylinders},}\ }\href@noop {} {\bibfield  {journal} {\bibinfo
  {journal} {Numer. Heat Transfer}\ }\textbf {\bibinfo {volume} {7(1)}},\
  \bibinfo {pages} {77--87} (\bibinfo {year} {1984})}\BibitemShut {NoStop}%
\bibitem [{\citenamefont {Lee}(1992)}]{lee2}%
  \BibitemOpen
  \bibfield  {author} {\bibinfo {author} {\bibfnamefont {T.~S.}\ \bibnamefont
  {Lee}},\ }\bibfield  {title} {\enquote {\bibinfo {title} {Laminar fluid
  convection between concentric and eccentric heated horizontal rotating
  cylinders for low-prandtl-number fluids},}\ }\href@noop {} {\bibfield
  {journal} {\bibinfo  {journal} {Int. J. Numer. Methods Fluids}\ }\textbf
  {\bibinfo {volume} {14}},\ \bibinfo {pages} {1037--1062} (\bibinfo {year}
  {1992})}\BibitemShut {NoStop}%
\bibitem [{\citenamefont {Gardiner}\ and\ \citenamefont
  {Sabersky}(1978)}]{gardiner}%
  \BibitemOpen
  \bibfield  {author} {\bibinfo {author} {\bibfnamefont {S.~R.~M.}\
  \bibnamefont {Gardiner}}\ and\ \bibinfo {author} {\bibfnamefont {R.~H.}\
  \bibnamefont {Sabersky}},\ }\bibfield  {title} {\enquote {\bibinfo {title}
  {Heat transfer in an annular gap},}\ }\href@noop {} {\bibfield  {journal}
  {\bibinfo  {journal} {Int. J. Heat Mass Transfer}\ }\textbf {\bibinfo
  {volume} {21}},\ \bibinfo {pages} {1459--1466} (\bibinfo {year}
  {1978})}\BibitemShut {NoStop}%
\bibitem [{\citenamefont {Childs}\ and\ \citenamefont
  {Long}(1996)}]{childs1996review}%
  \BibitemOpen
  \bibfield  {author} {\bibinfo {author} {\bibfnamefont {PRN}\ \bibnamefont
  {Childs}}\ and\ \bibinfo {author} {\bibfnamefont {CA}~\bibnamefont {Long}},\
  }\bibfield  {title} {\enquote {\bibinfo {title} {A review of forced
  convective heat transfer in stationary and rotating annuli},}\ }\href@noop {}
  {\bibfield  {journal} {\bibinfo  {journal} {Proc. Ins. Mech. Eng., Part C: J.
  Mech. Eng. Sci.}\ }\textbf {\bibinfo {volume} {210}},\ \bibinfo {pages}
  {123--134} (\bibinfo {year} {1996})}\BibitemShut {NoStop}%
\bibitem [{\citenamefont {Fusegi}\ \emph {et~al.}(1986)\citenamefont {Fusegi},
  \citenamefont {Farouk},\ and\ \citenamefont {Ball}}]{fusegi}%
  \BibitemOpen
  \bibfield  {author} {\bibinfo {author} {\bibfnamefont {T.}~\bibnamefont
  {Fusegi}}, \bibinfo {author} {\bibfnamefont {B.}~\bibnamefont {Farouk}}, \
  and\ \bibinfo {author} {\bibfnamefont {K.~S.}\ \bibnamefont {Ball}},\
  }\bibfield  {title} {\enquote {\bibinfo {title} {Mixed-convection flows
  within a horizontal concentric annulus with a heated rotating inner
  cylinder},}\ }\href@noop {} {\bibfield  {journal} {\bibinfo  {journal}
  {Numer. Heat Transfer}\ }\textbf {\bibinfo {volume} {9}},\ \bibinfo {pages}
  {591--604} (\bibinfo {year} {1986})}\BibitemShut {NoStop}%
\bibitem [{\citenamefont {Yoo}(1998)}]{yoo}%
  \BibitemOpen
  \bibfield  {author} {\bibinfo {author} {\bibfnamefont {J-S}\ \bibnamefont
  {Yoo}},\ }\bibfield  {title} {\enquote {\bibinfo {title} {Mixed-convection of
  air between two horizontal concentric cylinders with a cooled rotating outer
  cylinder},}\ }\href@noop {} {\bibfield  {journal} {\bibinfo  {journal} {Int.
  J. Heat Mass Transfer}\ }\textbf {\bibinfo {volume} {41(2)}},\ \bibinfo
  {pages} {293--302} (\bibinfo {year} {1998})}\BibitemShut {NoStop}%
\bibitem [{\citenamefont {Yang}\ and\ \citenamefont
  {Farouk}(1992)}]{yang1992three}%
  \BibitemOpen
  \bibfield  {author} {\bibinfo {author} {\bibfnamefont {Lei}\ \bibnamefont
  {Yang}}\ and\ \bibinfo {author} {\bibfnamefont {Bakhtier}\ \bibnamefont
  {Farouk}},\ }\bibfield  {title} {\enquote {\bibinfo {title}
  {Three-dimensional mixed convection flows in a horizontal annulus with a
  heated rotating inner circular cylinder},}\ }\href@noop {} {\bibfield
  {journal} {\bibinfo  {journal} {Int. J. Heat Mass Transfer}\ }\textbf
  {\bibinfo {volume} {35}},\ \bibinfo {pages} {1947--1956} (\bibinfo {year}
  {1992})}\BibitemShut {NoStop}%
\bibitem [{\citenamefont {Kahveci}(2016)}]{kahveci2016stability}%
  \BibitemOpen
  \bibfield  {author} {\bibinfo {author} {\bibfnamefont {K}~\bibnamefont
  {Kahveci}},\ }\bibfield  {title} {\enquote {\bibinfo {title} {Stability of
  unsteady mixed convection in a horizontal concentric annulus},}\ }\href@noop
  {} {\bibfield  {journal} {\bibinfo  {journal} {J. Appl. Fluid Mech.}\
  }\textbf {\bibinfo {volume} {9}},\ \bibinfo {pages} {2141--2147} (\bibinfo
  {year} {2016})}\BibitemShut {NoStop}%
\bibitem [{\citenamefont {Chhabra}\ and\ \citenamefont
  {Richardson}(2011)}]{chhabra}%
  \BibitemOpen
  \bibfield  {author} {\bibinfo {author} {\bibfnamefont {R.~P.}\ \bibnamefont
  {Chhabra}}\ and\ \bibinfo {author} {\bibfnamefont {J.~F.}\ \bibnamefont
  {Richardson}},\ }\enquote {\bibinfo {title} {Non-newtonian {F}low and
  {A}pplied {R}heology: {E}ngineering {A}pplications},}\ \ (\bibinfo
  {publisher} {Butterworth-{H}einemann},\ \bibinfo {year} {2011})\BibitemShut
  {NoStop}%
\bibitem [{\citenamefont {Morrison}(2001)}]{morrison}%
  \BibitemOpen
  \bibfield  {author} {\bibinfo {author} {\bibfnamefont {F.~A.}\ \bibnamefont
  {Morrison}},\ }\enquote {\bibinfo {title} {Understanding {R}heology, volume
  1},}\ \ (\bibinfo  {publisher} {Oxford {U}niversity {P}ress {N}ew {Y}ork},\
  \bibinfo {year} {2001})\BibitemShut {NoStop}%
\bibitem [{\citenamefont {Phan-Thien}\ and\ \citenamefont
  {Mai-Duy}(2013)}]{thien}%
  \BibitemOpen
  \bibfield  {author} {\bibinfo {author} {\bibfnamefont {N.}~\bibnamefont
  {Phan-Thien}}\ and\ \bibinfo {author} {\bibfnamefont {N.}~\bibnamefont
  {Mai-Duy}},\ }\enquote {\bibinfo {title} {Understanding {V}iscoelasticity:
  {A}n {I}ntroduction to {R}heology},}\ \ (\bibinfo  {publisher} {Springer},\
  \bibinfo {year} {2013})\BibitemShut {NoStop}%
\bibitem [{\citenamefont {Keunings}(1986)}]{keunings}%
  \BibitemOpen
  \bibfield  {author} {\bibinfo {author} {\bibfnamefont {R.}~\bibnamefont
  {Keunings}},\ }\bibfield  {title} {\enquote {\bibinfo {title} {On the high
  weissenberg number problem},}\ }\href@noop {} {\bibfield  {journal} {\bibinfo
   {journal} {J. Non-Newtonian Fluid Mech.}\ }\textbf {\bibinfo {volume}
  {20}},\ \bibinfo {pages} {209--226} (\bibinfo {year} {1986})}\BibitemShut
  {NoStop}%
\bibitem [{\citenamefont {Fattal}\ and\ \citenamefont
  {Kupferman}(2004)}]{fattal1}%
  \BibitemOpen
  \bibfield  {author} {\bibinfo {author} {\bibfnamefont {R.}~\bibnamefont
  {Fattal}}\ and\ \bibinfo {author} {\bibfnamefont {R.}~\bibnamefont
  {Kupferman}},\ }\bibfield  {title} {\enquote {\bibinfo {title} {Constitutive
  laws for the matrix-logarithm of the conformation tensor},}\ }\href@noop {}
  {\bibfield  {journal} {\bibinfo  {journal} {J. Non-Newtonian Fluid Mech.}\
  }\textbf {\bibinfo {volume} {123 (2-3)}},\ \bibinfo {pages} {281--285}
  (\bibinfo {year} {2004})}\BibitemShut {NoStop}%
\bibitem [{\citenamefont {Larson}\ \emph {et~al.}(1990)\citenamefont {Larson},
  \citenamefont {Shaqfeh},\ and\ \citenamefont {Muller}}]{larson}%
  \BibitemOpen
  \bibfield  {author} {\bibinfo {author} {\bibfnamefont {R.~G.}\ \bibnamefont
  {Larson}}, \bibinfo {author} {\bibfnamefont {E.~S.~G.}\ \bibnamefont
  {Shaqfeh}}, \ and\ \bibinfo {author} {\bibfnamefont {S.~J.}\ \bibnamefont
  {Muller}},\ }\bibfield  {title} {\enquote {\bibinfo {title} {A purely elastic
  instability in taylor-couette flow},}\ }\href@noop {} {\bibfield  {journal}
  {\bibinfo  {journal} {J. Fluid Mech.}\ }\textbf {\bibinfo {volume} {218}},\
  \bibinfo {pages} {573--600} (\bibinfo {year} {1990})}\BibitemShut {NoStop}%
\bibitem [{\citenamefont {Pakdel}\ and\ \citenamefont
  {McKinley}(1996)}]{pakdel}%
  \BibitemOpen
  \bibfield  {author} {\bibinfo {author} {\bibfnamefont {P.}~\bibnamefont
  {Pakdel}}\ and\ \bibinfo {author} {\bibfnamefont {G.~H.}\ \bibnamefont
  {McKinley}},\ }\bibfield  {title} {\enquote {\bibinfo {title} {Elastic
  instability and curved streamlines},}\ }\href@noop {} {\bibfield  {journal}
  {\bibinfo  {journal} {Phys. Rev. Lett.}\ }\textbf {\bibinfo {volume} {77
  (12)}},\ \bibinfo {pages} {2459} (\bibinfo {year} {1996})}\BibitemShut
  {NoStop}%
\bibitem [{\citenamefont {Steinberg}(2021)}]{steinberg}%
  \BibitemOpen
  \bibfield  {author} {\bibinfo {author} {\bibfnamefont {V.}~\bibnamefont
  {Steinberg}},\ }\bibfield  {title} {\enquote {\bibinfo {title} {Elastic
  turbulence: an experimental view on inertialess random flow},}\ }\href@noop
  {} {\bibfield  {journal} {\bibinfo  {journal} {Annu. Rev. Fluid Mech.}\
  }\textbf {\bibinfo {volume} {53}},\ \bibinfo {pages} {27--58} (\bibinfo
  {year} {2021})}\BibitemShut {NoStop}%
\bibitem [{\citenamefont {Groisman}\ and\ \citenamefont
  {Steinberg}(2000)}]{groisman1}%
  \BibitemOpen
  \bibfield  {author} {\bibinfo {author} {\bibfnamefont {A.}~\bibnamefont
  {Groisman}}\ and\ \bibinfo {author} {\bibfnamefont {V.}~\bibnamefont
  {Steinberg}},\ }\bibfield  {title} {\enquote {\bibinfo {title} {Elastic
  turbulence in a polymer solution flow},}\ }\href@noop {} {\bibfield
  {journal} {\bibinfo  {journal} {Nature}\ }\textbf {\bibinfo {volume} {405
  (6782)}},\ \bibinfo {pages} {53--55} (\bibinfo {year} {2000})}\BibitemShut
  {NoStop}%
\bibitem [{\citenamefont {Groisman}\ and\ \citenamefont
  {Steinberg}(2004)}]{groisman2}%
  \BibitemOpen
  \bibfield  {author} {\bibinfo {author} {\bibfnamefont {A.}~\bibnamefont
  {Groisman}}\ and\ \bibinfo {author} {\bibfnamefont {V.}~\bibnamefont
  {Steinberg}},\ }\bibfield  {title} {\enquote {\bibinfo {title} {Elastic
  turbulence in curvilinear flows of polymer solution},}\ }\href@noop {}
  {\bibfield  {journal} {\bibinfo  {journal} {New J. Phys.}\ }\textbf {\bibinfo
  {volume} {6 (1)}},\ \bibinfo {pages} {29} (\bibinfo {year}
  {2004})}\BibitemShut {NoStop}%
\bibitem [{\citenamefont {Whalley}\ \emph {et~al.}(2015)\citenamefont
  {Whalley}, \citenamefont {Abed}, \citenamefont {Dennis},\ and\ \citenamefont
  {Poole}}]{whalley}%
  \BibitemOpen
  \bibfield  {author} {\bibinfo {author} {\bibfnamefont {R.~D.}\ \bibnamefont
  {Whalley}}, \bibinfo {author} {\bibfnamefont {W.~M.}\ \bibnamefont {Abed}},
  \bibinfo {author} {\bibfnamefont {D.~J.~C.}\ \bibnamefont {Dennis}}, \ and\
  \bibinfo {author} {\bibfnamefont {R.~J.}\ \bibnamefont {Poole}},\ }\bibfield
  {title} {\enquote {\bibinfo {title} {Enhancing heat transfer at the
  micro-scale using elastic turbulence},}\ }\href@noop {} {\bibfield  {journal}
  {\bibinfo  {journal} {Theor. Appl. Mech. Lett.}\ }\textbf {\bibinfo {volume}
  {5 (3)}},\ \bibinfo {pages} {103--106} (\bibinfo {year} {2015})}\BibitemShut
  {NoStop}%
\bibitem [{\citenamefont {Li}\ \emph {et~al.}(2017)\citenamefont {Li},
  \citenamefont {Zhang}, \citenamefont {Cheng}, \citenamefont {Li},
  \citenamefont {Li}, \citenamefont {Qian},\ and\ \citenamefont {Joo}}]{li}%
  \BibitemOpen
  \bibfield  {author} {\bibinfo {author} {\bibfnamefont {D.-Y.}\ \bibnamefont
  {Li}}, \bibinfo {author} {\bibfnamefont {H.}~\bibnamefont {Zhang}}, \bibinfo
  {author} {\bibfnamefont {J.-P.}\ \bibnamefont {Cheng}}, \bibinfo {author}
  {\bibfnamefont {X.-B.}\ \bibnamefont {Li}}, \bibinfo {author} {\bibfnamefont
  {F.-C.}\ \bibnamefont {Li}}, \bibinfo {author} {\bibfnamefont
  {S.}~\bibnamefont {Qian}}, \ and\ \bibinfo {author} {\bibfnamefont {S.W.}\
  \bibnamefont {Joo}},\ }\bibfield  {title} {\enquote {\bibinfo {title}
  {Numerical simulation of heat transfer enhancement by elastic turbulence in a
  curvy microchannel},}\ }\href@noop {} {\bibfield  {journal} {\bibinfo
  {journal} {Microfluid Nanofluidics}\ }\textbf {\bibinfo {volume} {21 (2)}},\
  \bibinfo {pages} {25} (\bibinfo {year} {2017})}\BibitemShut {NoStop}%
\bibitem [{\citenamefont {Traore}\ \emph {et~al.}(2015)\citenamefont {Traore},
  \citenamefont {Castelain},\ and\ \citenamefont {Burghelea}}]{traore}%
  \BibitemOpen
  \bibfield  {author} {\bibinfo {author} {\bibfnamefont {B.}~\bibnamefont
  {Traore}}, \bibinfo {author} {\bibfnamefont {C.}~\bibnamefont {Castelain}}, \
  and\ \bibinfo {author} {\bibfnamefont {T.}~\bibnamefont {Burghelea}},\
  }\bibfield  {title} {\enquote {\bibinfo {title} {Efficient heat transfer in a
  regime of elastic turbulence},}\ }\href@noop {} {\bibfield  {journal}
  {\bibinfo  {journal} {J. Non-Newtonian Fluid Mech.}\ }\textbf {\bibinfo
  {volume} {223}},\ \bibinfo {pages} {62–76} (\bibinfo {year}
  {2015})}\BibitemShut {NoStop}%
\bibitem [{\citenamefont {Yao}\ \emph {et~al.}(2020)\citenamefont {Yao},
  \citenamefont {Yang}, \citenamefont {Zhao},\ and\ \citenamefont {Wen}}]{yao}%
  \BibitemOpen
  \bibfield  {author} {\bibinfo {author} {\bibfnamefont {G.}~\bibnamefont
  {Yao}}, \bibinfo {author} {\bibfnamefont {H.}~\bibnamefont {Yang}}, \bibinfo
  {author} {\bibfnamefont {J.}~\bibnamefont {Zhao}}, \ and\ \bibinfo {author}
  {\bibfnamefont {D.}~\bibnamefont {Wen}},\ }\bibfield  {title} {\enquote
  {\bibinfo {title} {Experimental study on flow and heat transfer enhancement
  by elastic instability in swirling flow},}\ }\href@noop {} {\bibfield
  {journal} {\bibinfo  {journal} {Int. J. Therm. Sci.}\ }\textbf {\bibinfo
  {volume} {157}},\ \bibinfo {pages} {106504} (\bibinfo {year}
  {2020})}\BibitemShut {NoStop}%
\bibitem [{\citenamefont {Poole}\ \emph {et~al.}(2012)\citenamefont {Poole},
  \citenamefont {Budhiraja}, \citenamefont {Cain},\ and\ \citenamefont
  {Scott}}]{poole}%
  \BibitemOpen
  \bibfield  {author} {\bibinfo {author} {\bibfnamefont {R.~J.}\ \bibnamefont
  {Poole}}, \bibinfo {author} {\bibfnamefont {B.}~\bibnamefont {Budhiraja}},
  \bibinfo {author} {\bibfnamefont {A.~R.}\ \bibnamefont {Cain}}, \ and\
  \bibinfo {author} {\bibfnamefont {P.~A.}\ \bibnamefont {Scott}},\ }\bibfield
  {title} {\enquote {\bibinfo {title} {Emulsification using elastic
  turbulence},}\ }\href@noop {} {\bibfield  {journal} {\bibinfo  {journal} {J.
  Non-Newtonian Fluid Mech.}\ }\textbf {\bibinfo {volume} {177}},\ \bibinfo
  {pages} {15--18} (\bibinfo {year} {2012})}\BibitemShut {NoStop}%
\bibitem [{\citenamefont {Burghelea}\ \emph {et~al.}(2004)\citenamefont
  {Burghelea}, \citenamefont {Segre},\ and\ \citenamefont
  {Steinberg}}]{burghelea2004mixing}%
  \BibitemOpen
  \bibfield  {author} {\bibinfo {author} {\bibfnamefont {T}~\bibnamefont
  {Burghelea}}, \bibinfo {author} {\bibfnamefont {Enrico}\ \bibnamefont
  {Segre}}, \ and\ \bibinfo {author} {\bibfnamefont {Victor}\ \bibnamefont
  {Steinberg}},\ }\bibfield  {title} {\enquote {\bibinfo {title} {Mixing by
  polymers: Experimental test of decay regime of mixing},}\ }\href@noop {}
  {\bibfield  {journal} {\bibinfo  {journal} {Phys. Rev. Lett.}\ }\textbf
  {\bibinfo {volume} {92}},\ \bibinfo {pages} {164501} (\bibinfo {year}
  {2004})}\BibitemShut {NoStop}%
\bibitem [{\citenamefont {Groisman}\ and\ \citenamefont
  {Steinberg}(2001)}]{groisman2001efficient}%
  \BibitemOpen
  \bibfield  {author} {\bibinfo {author} {\bibfnamefont {Alexander}\
  \bibnamefont {Groisman}}\ and\ \bibinfo {author} {\bibfnamefont {Victor}\
  \bibnamefont {Steinberg}},\ }\bibfield  {title} {\enquote {\bibinfo {title}
  {Efficient mixing at low reynolds numbers using polymer additives},}\
  }\href@noop {} {\bibfield  {journal} {\bibinfo  {journal} {Nature}\ }\textbf
  {\bibinfo {volume} {410}},\ \bibinfo {pages} {905--908} (\bibinfo {year}
  {2001})}\BibitemShut {NoStop}%
\bibitem [{\citenamefont {Grilli}\ \emph {et~al.}(2013)\citenamefont {Grilli},
  \citenamefont {V{\'a}zquez-Quesada},\ and\ \citenamefont
  {Ellero}}]{grilli2013transition}%
  \BibitemOpen
  \bibfield  {author} {\bibinfo {author} {\bibfnamefont {Muzio}\ \bibnamefont
  {Grilli}}, \bibinfo {author} {\bibfnamefont {Adolfo}\ \bibnamefont
  {V{\'a}zquez-Quesada}}, \ and\ \bibinfo {author} {\bibfnamefont {Marco}\
  \bibnamefont {Ellero}},\ }\bibfield  {title} {\enquote {\bibinfo {title}
  {Transition to turbulence and mixing in a viscoelastic fluid flowing inside a
  channel with a periodic array of cylindrical obstacles},}\ }\href@noop {}
  {\bibfield  {journal} {\bibinfo  {journal} {Phys. Rev. Lett.}\ }\textbf
  {\bibinfo {volume} {110}},\ \bibinfo {pages} {174501} (\bibinfo {year}
  {2013})}\BibitemShut {NoStop}%
\bibitem [{\citenamefont {Sasmal}(2023)}]{sasmal2023applications}%
  \BibitemOpen
  \bibfield  {author} {\bibinfo {author} {\bibfnamefont {C}~\bibnamefont
  {Sasmal}},\ }\bibfield  {title} {\enquote {\bibinfo {title} {Applications of
  elastic instability and elastic turbulence: Review, limitations, and future
  directions},}\ }\href@noop {} {\bibfield  {journal} {\bibinfo  {journal}
  {arXiv preprint arXiv:2301.02395}\ } (\bibinfo {year} {2023})}\BibitemShut
  {NoStop}%
\bibitem [{\citenamefont {Gupta}\ \emph {et~al.}(2022)\citenamefont {Gupta},
  \citenamefont {Chauhan},\ and\ \citenamefont {Sasmal}}]{gupta}%
  \BibitemOpen
  \bibfield  {author} {\bibinfo {author} {\bibfnamefont {S.}~\bibnamefont
  {Gupta}}, \bibinfo {author} {\bibfnamefont {A.}~\bibnamefont {Chauhan}}, \
  and\ \bibinfo {author} {\bibfnamefont {C.}~\bibnamefont {Sasmal}},\
  }\bibfield  {title} {\enquote {\bibinfo {title} {Influence of elastic
  instability and elastic turbulence on mixed convection of viscoelastic fluids
  in a lid-driven cavity},}\ }\href@noop {} {\bibfield  {journal} {\bibinfo
  {journal} {Int. J. Heat Mass Transfer}\ }\textbf {\bibinfo {volume} {186}},\
  \bibinfo {pages} {122469} (\bibinfo {year} {2022})}\BibitemShut {NoStop}%
\bibitem [{\citenamefont {Gupta}\ and\ \citenamefont
  {Sasmal}(2023)}]{gupta2023effect}%
  \BibitemOpen
  \bibfield  {author} {\bibinfo {author} {\bibfnamefont {S}~\bibnamefont
  {Gupta}}\ and\ \bibinfo {author} {\bibfnamefont {C}~\bibnamefont {Sasmal}},\
  }\bibfield  {title} {\enquote {\bibinfo {title} {Effect of cavity aspect
  ratio on mixed convective heat transfer phenomenon inside a lid-driven cavity
  due to elastic turbulence},}\ }\href@noop {} {\bibfield  {journal} {\bibinfo
  {journal} {Phys. Fluids}\ }\textbf {\bibinfo {volume} {35}} (\bibinfo {year}
  {2023})}\BibitemShut {NoStop}%
\bibitem [{\citenamefont {Cheng}\ \emph {et~al.}(2017)\citenamefont {Cheng},
  \citenamefont {Zhang}, \citenamefont {Cai}, \citenamefont {Li},\ and\
  \citenamefont {Li}}]{cheng2017effect}%
  \BibitemOpen
  \bibfield  {author} {\bibinfo {author} {\bibfnamefont {Jian-Ping}\
  \bibnamefont {Cheng}}, \bibinfo {author} {\bibfnamefont {Hong-Na}\
  \bibnamefont {Zhang}}, \bibinfo {author} {\bibfnamefont {Wei-Hua}\
  \bibnamefont {Cai}}, \bibinfo {author} {\bibfnamefont {Si-Ning}\ \bibnamefont
  {Li}}, \ and\ \bibinfo {author} {\bibfnamefont {Feng-Chen}\ \bibnamefont
  {Li}},\ }\bibfield  {title} {\enquote {\bibinfo {title} {Effect of polymer
  additives on heat transport and large-scale circulation in turbulent
  rayleigh-b{\'e}nard convection},}\ }\href@noop {} {\bibfield  {journal}
  {\bibinfo  {journal} {Phys. Rev. E}\ }\textbf {\bibinfo {volume} {96}},\
  \bibinfo {pages} {013111} (\bibinfo {year} {2017})}\BibitemShut {NoStop}%
\bibitem [{\citenamefont {Wei}\ \emph {et~al.}(2012)\citenamefont {Wei},
  \citenamefont {Ni},\ and\ \citenamefont {Xia}}]{wei2012enhanced}%
  \BibitemOpen
  \bibfield  {author} {\bibinfo {author} {\bibfnamefont {Ping}\ \bibnamefont
  {Wei}}, \bibinfo {author} {\bibfnamefont {Rui}\ \bibnamefont {Ni}}, \ and\
  \bibinfo {author} {\bibfnamefont {Ke-Qing}\ \bibnamefont {Xia}},\ }\bibfield
  {title} {\enquote {\bibinfo {title} {Enhanced and reduced heat transport in
  turbulent thermal convection with polymer additives},}\ }\href@noop {}
  {\bibfield  {journal} {\bibinfo  {journal} {Phys. Rev. E}\ }\textbf {\bibinfo
  {volume} {86}},\ \bibinfo {pages} {016325} (\bibinfo {year}
  {2012})}\BibitemShut {NoStop}%
\bibitem [{\citenamefont {Demir}(2003)}]{demir2003rayleigh}%
  \BibitemOpen
  \bibfield  {author} {\bibinfo {author} {\bibfnamefont {H}~\bibnamefont
  {Demir}},\ }\bibfield  {title} {\enquote {\bibinfo {title} {Rayleigh--benard
  convection of viscoelastic fluid},}\ }\href@noop {} {\bibfield  {journal}
  {\bibinfo  {journal} {Appl. Math. Comp.}\ }\textbf {\bibinfo {volume}
  {136}},\ \bibinfo {pages} {251--267} (\bibinfo {year} {2003})}\BibitemShut
  {NoStop}%
\bibitem [{\citenamefont {Cai}\ \emph {et~al.}(2019)\citenamefont {Cai},
  \citenamefont {Wei}, \citenamefont {Tang}, \citenamefont {Liu}, \citenamefont
  {Li},\ and\ \citenamefont {Li}}]{cai2019polymer}%
  \BibitemOpen
  \bibfield  {author} {\bibinfo {author} {\bibfnamefont {Weihua}\ \bibnamefont
  {Cai}}, \bibinfo {author} {\bibfnamefont {Tongzhou}\ \bibnamefont {Wei}},
  \bibinfo {author} {\bibfnamefont {Xiaojing}\ \bibnamefont {Tang}}, \bibinfo
  {author} {\bibfnamefont {Yao}\ \bibnamefont {Liu}}, \bibinfo {author}
  {\bibfnamefont {Biao}\ \bibnamefont {Li}}, \ and\ \bibinfo {author}
  {\bibfnamefont {Fengchen}\ \bibnamefont {Li}},\ }\bibfield  {title} {\enquote
  {\bibinfo {title} {The polymer effect on turbulent rayleigh-b{\'e}nard
  convection based on piv experiments},}\ }\href@noop {} {\bibfield  {journal}
  {\bibinfo  {journal} {Exp. Thermal Fluid Sci.}\ }\textbf {\bibinfo {volume}
  {103}},\ \bibinfo {pages} {214--221} (\bibinfo {year} {2019})}\BibitemShut
  {NoStop}%
\bibitem [{ope()}]{openfoam}%
  \BibitemOpen
  \href@noop {} {}\bibinfo {howpublished}
  {\url{https://www.openfoam.org}}\BibitemShut {NoStop}%
\bibitem [{\citenamefont {Pimenta}\ and\ \citenamefont
  {Alves}(2016)}]{rheotool}%
  \BibitemOpen
  \bibfield  {author} {\bibinfo {author} {\bibfnamefont {F.}~\bibnamefont
  {Pimenta}}\ and\ \bibinfo {author} {\bibfnamefont {M.}~\bibnamefont
  {Alves}},\ }\href@noop {} {\enquote {\bibinfo {title} {rheotool},}\ }\bibinfo
  {howpublished} {\url{https://github.com/fppimenta/rheoTool}} (\bibinfo {year}
  {2016})\BibitemShut {NoStop}%
\bibitem [{\citenamefont {Bejan}(2013)}]{bejan}%
  \BibitemOpen
  \bibfield  {author} {\bibinfo {author} {\bibfnamefont {A.}~\bibnamefont
  {Bejan}},\ }\enquote {\bibinfo {title} {Convection {H}eat {T}ransfer},}\ \
  (\bibinfo  {publisher} {John Wiley \& Sons},\ \bibinfo {year}
  {2013})\BibitemShut {NoStop}%
\bibitem [{\citenamefont {Bird}\ \emph {et~al.}(1987)\citenamefont {Bird},
  \citenamefont {Curtiss}, \citenamefont {Armstrong},\ and\ \citenamefont
  {Hassager}}]{bird}%
  \BibitemOpen
  \bibfield  {author} {\bibinfo {author} {\bibfnamefont {R.~B.}\ \bibnamefont
  {Bird}}, \bibinfo {author} {\bibfnamefont {C.~F.}\ \bibnamefont {Curtiss}},
  \bibinfo {author} {\bibfnamefont {R.~C.}\ \bibnamefont {Armstrong}}, \ and\
  \bibinfo {author} {\bibfnamefont {O.}~\bibnamefont {Hassager}},\ }\enquote
  {\bibinfo {title} {Dynamics of {P}olymeric {L}iquids, {V}olume 2: {K}inetic
  {T}heory},}\ \ (\bibinfo  {publisher} {Wiley-Interscience},\ \bibinfo {year}
  {1987})\BibitemShut {NoStop}%
\bibitem [{\citenamefont {James}(2009)}]{james}%
  \BibitemOpen
  \bibfield  {author} {\bibinfo {author} {\bibfnamefont {D.~F.}\ \bibnamefont
  {James}},\ }\bibfield  {title} {\enquote {\bibinfo {title} {Boger fluids},}\
  }\href@noop {} {\bibfield  {journal} {\bibinfo  {journal} {Annu. Rev. Fluid
  Mech.}\ }\textbf {\bibinfo {volume} {41}},\ \bibinfo {pages} {129--142}
  (\bibinfo {year} {2009})}\BibitemShut {NoStop}%
\bibitem [{\citenamefont {Alves}\ \emph {et~al.}(2003)\citenamefont {Alves},
  \citenamefont {Oliveira},\ and\ \citenamefont {Pinho}}]{cubista}%
  \BibitemOpen
  \bibfield  {author} {\bibinfo {author} {\bibfnamefont {M.~A.}\ \bibnamefont
  {Alves}}, \bibinfo {author} {\bibfnamefont {P.~J.}\ \bibnamefont {Oliveira}},
  \ and\ \bibinfo {author} {\bibfnamefont {F.~T.}\ \bibnamefont {Pinho}},\
  }\bibfield  {title} {\enquote {\bibinfo {title} {A convergent and universally
  bounded interpolation scheme for the treatment of advection},}\ }\href@noop
  {} {\bibfield  {journal} {\bibinfo  {journal} {Int. J. Numer. Methods
  Fluids}\ }\textbf {\bibinfo {volume} {41}},\ \bibinfo {pages} {47--75}
  (\bibinfo {year} {2003})}\BibitemShut {NoStop}%
\bibitem [{\citenamefont {Ajiz}\ and\ \citenamefont {Jennings}(1984)}]{DIC}%
  \BibitemOpen
  \bibfield  {author} {\bibinfo {author} {\bibfnamefont {M.~A.}\ \bibnamefont
  {Ajiz}}\ and\ \bibinfo {author} {\bibfnamefont {A.}~\bibnamefont
  {Jennings}},\ }\bibfield  {title} {\enquote {\bibinfo {title} {A robust
  incomplete choleski-conjugate gradient algorithm},}\ }\href@noop {}
  {\bibfield  {journal} {\bibinfo  {journal} {Int. J. Numer. Methods Eng.}\
  }\textbf {\bibinfo {volume} {20 (5)}},\ \bibinfo {pages} {949--966} (\bibinfo
  {year} {1984})}\BibitemShut {NoStop}%
\bibitem [{\citenamefont {Lee}\ \emph {et~al.}(2003)\citenamefont {Lee},
  \citenamefont {Zhang},\ and\ \citenamefont {Lu}}]{DILU}%
  \BibitemOpen
  \bibfield  {author} {\bibinfo {author} {\bibfnamefont {J.}~\bibnamefont
  {Lee}}, \bibinfo {author} {\bibfnamefont {J.}~\bibnamefont {Zhang}}, \ and\
  \bibinfo {author} {\bibfnamefont {C.~C.}\ \bibnamefont {Lu}},\ }\bibfield
  {title} {\enquote {\bibinfo {title} {Incomplete lu preconditioning for large
  scale dense complex linear systems from electromagnetic wave scattering
  problems},}\ }\href@noop {} {\bibfield  {journal} {\bibinfo  {journal} {J.
  Comput. Phys.}\ }\textbf {\bibinfo {volume} {185 (1)}},\ \bibinfo {pages}
  {158--175} (\bibinfo {year} {2003})}\BibitemShut {NoStop}%
\bibitem [{\citenamefont {Fattal}\ and\ \citenamefont
  {Kupferman}(2005)}]{fattal2}%
  \BibitemOpen
  \bibfield  {author} {\bibinfo {author} {\bibfnamefont {R.}~\bibnamefont
  {Fattal}}\ and\ \bibinfo {author} {\bibfnamefont {R.}~\bibnamefont
  {Kupferman}},\ }\bibfield  {title} {\enquote {\bibinfo {title}
  {Time-dependent simulation of viscoelastic flows at high weissenberg number
  using the log-conformation representation},}\ }\href@noop {} {\bibfield
  {journal} {\bibinfo  {journal} {J. Non-Newtonian Fluid Mech.}\ }\textbf
  {\bibinfo {volume} {126}},\ \bibinfo {pages} {23--37} (\bibinfo {year}
  {2005})}\BibitemShut {NoStop}%
\bibitem [{\citenamefont {Pimenta}\ and\ \citenamefont
  {Alves}(2017)}]{pimenta}%
  \BibitemOpen
  \bibfield  {author} {\bibinfo {author} {\bibfnamefont {F.}~\bibnamefont
  {Pimenta}}\ and\ \bibinfo {author} {\bibfnamefont {M.~A.}\ \bibnamefont
  {Alves}},\ }\bibfield  {title} {\enquote {\bibinfo {title} {Stabilization of
  an open-source finite-volume solver for viscoelastic fluid flows},}\
  }\href@noop {} {\bibfield  {journal} {\bibinfo  {journal} {J. Non-Newtonian
  Fluid Mech.}\ }\textbf {\bibinfo {volume} {239}},\ \bibinfo {pages} {85--104}
  (\bibinfo {year} {2017})}\BibitemShut {NoStop}%
\bibitem [{\citenamefont {Zheng}\ \emph {et~al.}(2023)\citenamefont {Zheng},
  \citenamefont {Boutaous}, \citenamefont {Xin}, \citenamefont {Siginer},\ and\
  \citenamefont {Cai}}]{zheng2023time}%
  \BibitemOpen
  \bibfield  {author} {\bibinfo {author} {\bibfnamefont {Xin}\ \bibnamefont
  {Zheng}}, \bibinfo {author} {\bibfnamefont {M'hamed}\ \bibnamefont
  {Boutaous}}, \bibinfo {author} {\bibfnamefont {Shihe}\ \bibnamefont {Xin}},
  \bibinfo {author} {\bibfnamefont {Dennis~A}\ \bibnamefont {Siginer}}, \ and\
  \bibinfo {author} {\bibfnamefont {Weihua}\ \bibnamefont {Cai}},\ }\bibfield
  {title} {\enquote {\bibinfo {title} {Time-dependent oscillating viscoelastic
  rayleigh-b{\'e}nard convection: Viscoelastic kinetic energy budget
  analysis},}\ }\href@noop {} {\bibfield  {journal} {\bibinfo  {journal} {Phys.
  Rev. Fluids}\ }\textbf {\bibinfo {volume} {8}},\ \bibinfo {pages} {023303}
  (\bibinfo {year} {2023})}\BibitemShut {NoStop}%
\bibitem [{\citenamefont {James}\ and\ \citenamefont
  {Tripathi}(2023)}]{james2023pressure}%
  \BibitemOpen
  \bibfield  {author} {\bibinfo {author} {\bibfnamefont {David~F}\ \bibnamefont
  {James}}\ and\ \bibinfo {author} {\bibfnamefont {Abhishek}\ \bibnamefont
  {Tripathi}},\ }\bibfield  {title} {\enquote {\bibinfo {title} {Pressure drop
  in a converging channel with viscoelastic polymer solutions having power-law
  viscous behaviour},}\ }\href@noop {} {\bibfield  {journal} {\bibinfo
  {journal} {J. Non-Newtonian Fluid Mech.}\ }\textbf {\bibinfo {volume}
  {312}},\ \bibinfo {pages} {104974} (\bibinfo {year} {2023})}\BibitemShut
  {NoStop}%
\bibitem [{\citenamefont {Bendov{\'a}}\ \emph {et~al.}(2009)\citenamefont
  {Bendov{\'a}}, \citenamefont {{\v{S}}i{\v{s}}ka},\ and\ \citenamefont
  {Macha{\v{c}}}}]{bendova2009pressure}%
  \BibitemOpen
  \bibfield  {author} {\bibinfo {author} {\bibfnamefont {Helena}\ \bibnamefont
  {Bendov{\'a}}}, \bibinfo {author} {\bibfnamefont {Bed{\v{r}}ich}\
  \bibnamefont {{\v{S}}i{\v{s}}ka}}, \ and\ \bibinfo {author} {\bibfnamefont
  {Ivan}\ \bibnamefont {Macha{\v{c}}}},\ }\bibfield  {title} {\enquote
  {\bibinfo {title} {Pressure drop excess in the flow of viscoelastic liquids
  through fixed beds of particles},}\ }\href@noop {} {\bibfield  {journal}
  {\bibinfo  {journal} {Chem. Eng. Proc. : Process Intensification}\ }\textbf
  {\bibinfo {volume} {48}},\ \bibinfo {pages} {29--37} (\bibinfo {year}
  {2009})}\BibitemShut {NoStop}%
\end{thebibliography}%

\end{document}